\documentclass[aps,pra,twocolumn,superscriptaddress,amsmath,amssymb]{revtex4-2}
\usepackage{graphicx}
\usepackage{amsmath, amssymb, bm, physics, mathtools}
\usepackage{tcolorbox}
\usepackage{xcolor}
\usepackage{multirow}
\usepackage{array}
\usepackage{booktabs}
\usepackage{braket}
\usepackage{enumitem}
\usepackage{float}
\usepackage{placeins}
\usepackage{appendix}

\newcommand{\Hil}{H_{\rm cb}}
\newcommand{\HilT}{\tilde{H}_{\rm cb}}
\newcommand{\Hbath}{H_{\rm bath}}
\newcommand{\HcbB}{H_{\rm cbB}}
\newcommand{\HcbBT}{\tilde{H}_{\rm cbB}}
\newcommand{\Lc}{\mathcal{L}}
\newcommand{\Cc}{\mathcal{C}}
\newcommand{\one}{\openone}              
\newcommand{\sgp}[1]{\sigma_+^{#1}}
\newcommand{\sgm}[1]{\sigma_-^{#1}}
\newcommand{\sgz}[1]{\sigma_z^{#1}}

\newcommand{\nF}{n_{\rm F}}
\newcommand{\nS}{n_{\rm S}}
\newcommand{\nD}{n_{\rm D}}

\newcommand{\pgg}{\ket{\psi_{11}}}          
\newcommand{\pee}{\ket{\psi_{00}}}            
\newcommand{\ppl}{\ket{\psi_{01}}}            
\newcommand{\pmi}{\ket{\psi_{10}}}

\begin{document}

\title{Two-qubit charger-battery system subject to weak, continuous measurements with quantum point contacts}

\author{Banumathy M}
\affiliation{Department of Physics, National Institute of Technology Calicut \\ Kozhikode, Kerala, India 673601}
\affiliation{School of Physics, Indian Institute of Science Education and Research Thiruvananthapuram \\ Thiruvananthapuram, Kerala, India 695551}

\author{Anil Shaji}
\affiliation{School of Physics, Indian Institute of Science Education and Research Thiruvananthapuram \\ Thiruvananthapuram, Kerala, India 695551}
\affiliation{Centre for High Performance Computing, Indian Institute of Science Education and Research Thiruvananthapuram \\ Thiruvananthapuram, Kerala, India 695551}


\begin{abstract}
   Quantum batteries modeled as two-qubit systems coupled to Markovian thermal reservoirs have been shown to benefit from measurement-assisted charging, where projective measurements enhance the charging rate at an infinite thermodynamic resource cost. In this work, we consider weak, continuous measurements implemented via quantum point contact detectors (QPC), which enhance the charging rate at a definite and quantifiable resource cost. We analyze three measurement configurations namely, a single QPC, two independent QPCs, and a series-coupled (coherent) two-QPC scheme, and study their effect on the steady-state charging rate, defined as the rate of energy flow from the charger qubit to the battery qubit, relative to the unmeasured baseline. We find that the charging rate enhancement is non-monotonic as a function of the temperature gradient and potential gradient required to drive the QPCs, exhibiting a plateau of near-optimal enhancement. Comparing the three configurations, the plateau of optimal enhancement contracts toward lower temperature and chemical potential for both the cases with two QPCs compared to the single QPC case. The coherent measurement further shows a lowering of the resource requirement relative to the two independent QPC case for achieving the same enhancement.   The hierarchy is coherent greater than two independent QPCs which is is greater than single QPC with respect to both the magnitude of the charging rate enhancement and the minimization of measurement resources.
\end{abstract}

\maketitle

\section{Introduction}

Measurements in classical thermodynamics are passive acts from which information about the macro-state of the system of interest can be obtained when thermodynamic variables like temperature, pressure, etc.\ are measured~\cite{Maxwell1871,Szilard1929,LeffRex2003,Leff2014,Bennett1982}. The act of measurement, in principle, has no effect on the state of the system in this case, and the information gained cannot, per se, be used to extract useful work from the system~\cite{Maxwell1871,Szilard1929,LeffRex2003,Leff2014,Bennett1982}. Measurements in the classical context can also ascertain the micro-state of the system. Even though this information is obtained without disturbance to the system, unlike in the previous case, the information can be used to extract useful work as can be done by the proverbial Maxwell's demon~\cite{Maxwell1871,Szilard1929,LeffRex2003,Leff2014,Bennett1982}. There is only information exchange between the system and the demon as a result of the measurement and there is no energy exchange. The cost in terms of energy of the information gain and consequent reduction in entropy has to be squared off at the point where the information has to be erased, as pointed out by Landauer~\cite{Landauer1961,Bennett1982,Bennett2003}.

Measurements on quantum systems are not passive acts and they involve exchange of not only information but also energy, momentum, angular momentum, etc. The energy cost of information transfer and associated entropy changes are often accounted for during the act of measurement itself, and disturbances to the measured system may be unavoidable as well~\cite{Zurek1991,Busch1996,Peres1995,Jacobs2014,Landauer1961,Bennett2003,JacobsSteck2006,Ozawa2003}. Quantum measurements have been proven to supply energy into the system, either by direct projective measurement or via weak continuous measurement. The role of the measurement apparatus is crucial in this regard. The microscopic model of the apparatus matters, as it determines the nature of
the energy exchange.

In this work, we investigate measurement-powered quantum batteries under different measurement configurations. We consider a two-qubit system where one qubit acts as a charger and the other as a battery. The charger is connected to a hot reservoir which provides it with a supply of energy, while the battery is coupled to a cold reservoir which allows one to extract energy from it. The coupling between the charger and the battery allows for
{\em charging} the battery. The key figures of merit in the context of quantum batteries are the charging/discharging rates and ergotropy~\cite{Gherardini2020,Seah2021}, etc. Enhancement in the charging/discharging rates of a quantum battery relative to an equivalent classical battery, which can hold exactly the same amount of energy as the quantum one when fully charged, is often the focus in the contemporary
literature on quantum batteries~\cite{BuffoniQMC2019,Bhandari2023,Du2025,Yao2023}.

In Ref.~\cite{Elouard2025} a model of a thermal machine---specifically a refrigerator made of two qubits---coupled to a quantum point contact which performs weak, continuous measurements on the machine is considered in detail. Building on this work, we consider the two-qubit quantum battery and charger system, rather than a refrigerator, and systematically compare three measurement protocols: (i)~single-qubit measurements on the charger, (ii)~correlated two-qubit measurement on both qubits via series-coupled QPCs, and (iii)~independent two-qubit measurement via parallel QPCs. Our analysis reveals how measurement geometry affects charging rates, identifies optimal operating regimes, and demonstrates a quantum advantage from correlated measurements. Measurements have been shown previously to improve the charging rate and ergotropy in charger-battery systems~\cite{BuffoniQMC2019,Bhandari2023,ElouardJordan2018} but the question of how much energy is introduced into the system by the quantum measurement back-action has not been considered systematically. In the context of batteries this is a significant question, since it is important to clarify whether the charging rate is increased due to a direct injection of energy by the measurement or through more subtle quantum mechanical means. When considering projective measurements, the fact that such measurements have an infinite energy cost in thermodynamic terms~\cite{ElouardRole2017,Manikandan2018} has to be considered while evaluating the effectiveness of measurements in increasing the efficiency and speed of quantum batteries and other quantum thermal machines. By explicitly modeling the measurement apparatus as in Ref.~\cite{Elouard2025}, we are able to quantify the energy cost of performing the measurement as well as the energy injected into the system of interest by the measurement back-action.

The paper is organized as follows. Sec.~\ref{sec:system} introduces the two-qubit system and general framework. Sec.~\ref{sec:MeasurementDissipator} develops the single-QPC measurement model while Sec.~\ref{sec:correlated} extends it to  measurements using two independent and series-coupled QPCs.  A comparative discussion of the cases and our conclusions are in Sec.~\ref{sec:discussion}.

\section{The Two-Qubit Charger and Battery System}
\label{sec:system}

We consider a system composed of two charge qubits, each coupled to separate thermal electron reservoirs at different temperatures. For concreteness we treat each qubit as a double quantum dot with a single electron shared between them. The presence of the electron on one of the two dots corresponds to the state $|0\rangle$ and its presence in the other dot in the pair corresponds to the $|1\rangle$ state. The tunnel barrier between the two dots is configured so that the thermal reservoir connected to the qubit in question can drive transitions across it and place each qubit in states that are arbitrary superpositions of $|0\rangle$ and $|1\rangle$. The coupling between the qubits is assumed to be capacitive in nature and does not allow transfer of electrons from one qubit to another. Of the two qubits, the one on the left is taken to be the ``charger,'' as it is coupled to the reservoir at the higher temperature, while the one on the right is the ``battery,'' which is charged through its interaction with the charger qubit and discharged by the reservoir at lower temperature connected to it. An additional external drive may also be considered for the charger qubit but in the following we focus exclusively on the thermal driving of the battery and charger systems. 

\subsection{System Hamiltonian and Eigenstates}

Throughout this paper, we use natural units where $k_B = 1$ and $\hbar = 1$. The Hamiltonian of the two-qubit system is given by
\begin{equation}
\Hil = \frac{\omega_c}{2}\sgz{c}+\frac{\omega_b}{2}\sgz{b}
  +\gamma\!\left(\sgp{c}\sgm{b}+\sgm{c}\sgp{b}\right).
\label{eq:hamiltonian}
\end{equation}
where $\omega_c$ and $\omega_b$ represent the energy spacings of the charger and battery qubits, respectively. Here, $\sigma_z^i$ is the Pauli operator in the $z$-direction, and $\sigma_\pm^i$ denote the raising/lowering operators. The parameter $\gamma$ controls the interaction strength between the two qubits. A term of the form $\Omega_D\!\left(\sgp{c}e^{-i\omega_d t}+\sgm{c}e^{+i\omega_d t}\right)$ should be added if an external drive is to be considered but here we choose $\Omega_D=0$. We use the standard computational basis to denote the non-interacting states of the two qubits where $\ket{0}$ is the \emph{excited} state and $\ket{1}$ the \emph{ground} state, so that $\sgz{\alpha}\ket{0}_\alpha = +\ket{0}_\alpha$, $\sgz{\alpha}\ket{1}_\alpha = -\ket{1}_\alpha$, $\sgp{\alpha}\ket{1}_\alpha = \ket{0}_\alpha$ and, $\sgm{\alpha}\ket{0}_\alpha = \ket{1}_\alpha$. The two-qubit product basis $\{\ket{00},\ket{01},\ket{10},\ket{11}\}$ are the eigenstates on the non-interacting part of the Hamiltonian.  The eigenstates of the full interacting Hamiltonian are the \emph{dressed} states, denoted as $\pee,\ppl,\pmi,\pgg$.

The eigenvalues of $H_{\rm sys}$ are 
\begin{eqnarray*}
    \Lambda_{00} = -\Lambda_{11} & = &  (\omega_{\rm c} + \omega_{\rm b})/2, \\
    \Lambda_{01} & = &  \frac{1}{2} \sqrt{\Delta^2 + 4\gamma^2} = \Omega, \\
    \Lambda_{10} & = &  -\frac{1}{2} \sqrt{\Delta^2 + 4\gamma^2} = -\Omega,
\end{eqnarray*}  
where $\Delta = \omega_c - \omega_b$ is the  difference in energy spacings of the two qubits~\cite{Du2025} and $\Omega \equiv \sqrt{\Delta^2 + 4\gamma^2}/2$. The eigenstates of $\Hil$ can be written in terms of the eigenstates of the non-interacting Hamiltonian as, 
\begin{align}
    |\psi_{00}\rangle &= |00\rangle, \qquad |\psi_{11}\rangle = |11\rangle, \nonumber \\
    |\psi_{01}\rangle & = \cos\tfrac{\theta}{2}|01\rangle
                        + \sin\tfrac{\theta}{2}|10\rangle, \nonumber \\
    |\psi_{10}\rangle &= -\sin\tfrac{\theta}{2}|01\rangle
                        + \cos\tfrac{\theta}{2}|10\rangle, 
    \label{eigstates}
\end{align}
with the mixing angle
\begin{equation}
\theta = \sin^{-1}\!\!\left(\!\frac{2\gamma}{\sqrt{\Delta^2 + 4\gamma^2}}\!\right)\!\!
       =\! \sin^{-1}\!\!\left(\frac{\gamma}{\Omega}\right)\!= \tan^{-1}\!\!\left(\!\frac{2\gamma}{\Delta}\!\right)\!.
\label{eq:mixing_angle}
\end{equation}

We assume that there is a thermal reservoir of non-interacting Fermions on the left, at temperature $T_{\rm c}$, that couples to the charger qubit, and an independent Fermionic reservoir on the right at temperature $T_{\rm b} < T_{\rm c}$ that couples to the battery qubit. In the computational basis, the charger reservoir mediates the transitions $|11\rangle \leftrightarrow |01\rangle$ and $|10\rangle \leftrightarrow |00\rangle$, while the battery reservoir mediates $|01\rangle \leftrightarrow |00\rangle$ and $|10\rangle \leftrightarrow |11\rangle$. The full Hamiltonian of the two qubits plus the baths (denoted by the subscript B) is given by,
\begin{equation}
H_\text{tot}=\Hil+H_{\rm B}+H_\text{cbB},
\end{equation}
with
\begin{equation}
  H_{\rm B} =  \sum_{k}\epsilon_k^c f_k^{c\dagger} f_k^c + \sum_{k} \epsilon_k^b f_k^{b\dagger} f_k^b, \label{eq:Hbath} 
\end{equation}
and
\begin{eqnarray}
  H_\text{cbB} &=&\sum_{k}\lambda_k^c\!\left(\sgp{c}f_k^c+\sgm{c}f_k^{c\dagger}\right) \nonumber \\
  && \qquad \qquad + \sum_{k} \lambda_k^b \! \left( \sgp{b}f_k^b + \sgm{b}f_k^{b\dagger} \right).
  \label{eq:Hsb}
\end{eqnarray}
The fermionic operators satisfy $\{f_k^\alpha,f_{k'}^{\alpha'\dagger}\}=\delta_{kk'}\delta_{\alpha\alpha'}$ and we assume that the coupling constants $\lambda_k$ are real.
Each bath is in thermal equilibrium characterized by the Fermi-Dirac distribution:
\begin{equation}
  \langle f_k^{\alpha\dagger}f_k^\alpha\rangle_{\rm B}
  =\nF(\epsilon_k^\alpha,T_\alpha)=\! \frac{1}{e^{\epsilon_k^\alpha/T_\alpha}+1}, \; \alpha\in\{\rm{c}, \rm{b}\}.
\end{equation}
Here we have set the chemical potentials of the baths equal to zero. 

\subsection{Effect of the baths \label{baths1}}
We now proceed to characterize the effect of the baths on the charger-battery system in terms of jump operators that appear in the Gorini-Kossakowski-Sudarshan-Lindblad (GKSL) \cite{Gorini1976, Lindblad1976, Kossakowski1972} master equation describing the open evolution of the two qubits. The first step is to move to the interaction picture with respect to $H_0 = \Hil + \Hbath$ in which Schrodinger picture operators, $O$ acquire a time dependence, $\tilde{O}(t)=e^{iH_0 t}O\,e^{-iH_0 t}$. The interaction picture density matrix, $\tilde{\varrho}(t)$, of the total system consisting of the charger, battery and, the baths,
obeys the von~Neumann equation,
\begin{equation}
  \frac{d\tilde{\varrho}(t)}{dt}=-i\bigl[\HcbBT(t),\tilde{\varrho}(t)\bigr].
  \label{eq:vN}
\end{equation}
Formally integrating Eq.~\eqref{eq:vN} and substituting back, the exact equation of motion for the reduced system density matrix in the interaction picture $\tilde{\rho}(t) \equiv \tilde{\rho}_{\rm cb}(t) = \Tr_{\rm B} [\tilde{\varrho}(t)]$ is,
\begin{equation}
  \frac{d\rho(t)}{dt}=-\int_0^t ds\,\Tr_{\rm B} \Bigl[\HcbBT(t), \bigl[\HcbBT(s),\tilde{\varrho}(s)\bigr]\Bigr].
  \label{eq:exact}
\end{equation}
The first-order term, $-i\Tr_{\rm B}[\HcbBT(t), \tilde{\varrho}(0)]$ vanishes because $\langle f_k^\alpha\rangle_{\rm B}=\langle f_k^{\alpha \dagger} \rangle_{\rm B}=0$ for the equilibrium thermal state of the bath. 

To proceed further we need to make a sequence of three approximations. The first one is the Born approximation which is applicable if the system-bath coupling is weak ($\lambda_k^\alpha$ small), so the total state remains approximately a product state at all times of the form, 
\begin{equation}
  \tilde{\varrho}(t)\approx \tilde{\rho}(t) \otimes \rho_{\rm B}, \quad \rho_{\rm B} = \rho_\text{B}^c \otimes \rho_{\rm B}^b,
\end{equation}
where $\rho_{\rm B}^\alpha$ is the thermal equilibrium state of bath $\alpha$.
Under this approximation, Eq.~\eqref{eq:exact} becomes,
\begin{equation}
  \frac{d\tilde{\rho}(t)}{dt}\!= \!- \!\!\int_0^t \!\!\! ds\,\Tr_{\rm B} \Bigl[\HcbBT(t), \bigl[\HcbBT(s),\tilde{\rho}(s) \otimes \rho_{\rm B}\bigr]\Bigr],
  \label{eq:born}
\end{equation}
which is a second order equation in $\lambda_k^\alpha$. 
in the interaction picture, the Fermionic bath operators $f_k^\alpha$ acquires a time dependence of the form,
$\tilde{f}_k^\alpha(t)=f_k^\alpha e^{-i\epsilon_k^\alpha t}$. Denoting $\tilde{B}_\alpha(t)=\sum_k\lambda_k^\alpha f_k^\alpha e^{-i\epsilon_k^\alpha t}$, we see that the trace over the bath of the system-bath coupling terms in Eq.~\eqref{eq:born} produces bath correlation
functions. For bath $\alpha$ the relevant two-time correlation functions are,
\begin{eqnarray}
  \Cc_\alpha^+(t-s)&= &\Tr_{\rm B}\!\left[ \tilde{B}_\alpha^\dagger(t)\tilde{B}_\alpha(s)\,\rho_{\rm B}^\alpha\right] \nonumber \\ 
  &=& \sum_k (\lambda_k^\alpha)^2 \nF (\epsilon_k^\alpha )e^{i\epsilon_k^\alpha (t-s)},\label{eq:C+}\\
  \Cc_\alpha^-(t-s) & = & \Tr_{\rm B}\!\left[ \tilde{B}_\alpha(t) \tilde{B}_\alpha^\dagger(s) \, \rho_{\rm B}^\alpha \right] \nonumber \\
& = & \sum_k (\lambda_k^\alpha)^2 [1-\nF (\epsilon_k^\alpha) ]e^{-i\epsilon_k^\alpha (t-s)},\label{eq:C-}
\end{eqnarray}
The functions $\Cc_\alpha^+$ and $\Cc_\alpha^-$ encode absorption and emission processes respectively, weighted by the Fermi-Dirac distribution. $\Cc_\alpha^+$ corresponds to absorption of a quantum by the respective qubit with the bath emitting the quantum while $\Cc_\alpha^-$ corresponds to the reverse process where the qubit emits a quantum with the bath absorbs it. 

We now introduce the bath spectral density, 
\begin{equation}
J_\alpha(\epsilon) \equiv (\lambda^\alpha(\epsilon))^2 D_\alpha (\epsilon),
  \label{eq:spectral_density}
\end{equation}
where $D_\alpha(\epsilon)$ is the density of states of bath $\alpha$ at energy $\epsilon$ which counts the number of Fermionic modes per unit energy. Using the spectral density we can convert the sums in Eq.~\eqref{eq:C+} and Eq.~\eqref{eq:C-} into integrals as
\begin{eqnarray}
    \Cc_\alpha^+(\tau)&= & \int_{-\infty}^\infty d\epsilon\;J_\alpha(\epsilon) \nF(\epsilon,T_\alpha)\,e^{+i\epsilon\tau}, \nonumber \\
    \Cc_\alpha^-(\tau)&= & \int_{-\infty}^\infty d\epsilon\;J_\alpha(\epsilon) [1-\nF(\epsilon,T_\alpha)]\,e^{-i\epsilon\tau}.
\end{eqnarray}
If the bath spectral density is a very broad function of $\epsilon$ we see that the bath correlation functions decay very rapidly. This, in turn, is characteristic of large baths for which any small disturbance produced by its interaction with the respective qubits is wiped out almost instantaneously with the bath returning to its equilibrium thermal state very rapidly. The wide and flat spectral density also means that $J_\alpha(\epsilon) \sim J_\alpha = (\lambda^\alpha)^2 D_\alpha$. In other words the charger and bath qubits are coupled to all the modes of their respective baths via constant couplings, $\lambda^{\rm c}$ and $\lambda^{\rm b}$ respectively.  Assuming that the baths connected to the charger and battery qubits is of this kind with very short relaxation time-scales, $\tau_{\rm B}$, we can use the second approximation, namely assuming that the dynamics of the battery and charger is Markovian. 

There are two steps in applying the Markov approximation. The first one involves replacing $\tilde{\rho}_{\rm cb} (s) \to \tilde{\rho}_{\rm cb}(t)$ in the integrand of Eq.~\eqref{eq:born} since within the short bath relaxation timescale, the state $\tilde{\rho}_{\rm cb}$ has hardly changed. The second step involves extending the upper limit of the integral in Eq.~\eqref{eq:born} to infinity with $\tau = t-s$. This is justified because beyond the bath relaxation time-scale, the integrand is practically zero and terms with $\tau>\tau_{\rm B}$ contribute negligibly. Applying the Markov approximation we get the time-local Redfield master equation in the interaction picture,  
\begin{equation}
  \frac{d\tilde{\rho} (t)}{dt}\!\!=\!-\!\!\int_0^\infty \!\!\!\! d\tau \, Tr_{\rm B}
  \Bigl[\HcbBT(t),  \bigl[\HcbBT(t-\tau),\tilde{\rho}(t)\otimes\rho_{\rm B}\bigr]\Bigr].
  \label{eq:redfield0}
\end{equation}

The operators $\sigma^\alpha_{\pm}$ that appear in $\HcbB$ also acquire a time dependence in the interaction picture because $[\Hil, \sigma^\alpha_{\pm}] \neq 0$. In order to proceed further these operators have to be decomposed into eigen-operators (Fourier component) of $\Hil$ with definite frequencies. Any operator $X_\alpha$ can be decomposed into eigenoperators of $\Hil$ as $X_\alpha = \sum_\omega X_\alpha(\omega)$ where
\begin{equation} 
\label{eq:eigenops1}
X_\alpha(\omega) = \sum_{\substack{i,j;l,m:\\ \Lambda_{ij}-\Lambda_{lm}=\omega}} \langle \psi_{ij} |X_\alpha |\psi_{lm}\rangle \,|\psi_{ij}\rangle \langle \psi_{lm}|,
\end{equation}
where $\Lambda_{ij}$ and $|\psi_{ij}\rangle$, as previously introduced, are the eigenvalues and eigenstates of $\Hil$. By construction $[\Hil, X_\alpha(\omega)] = -\omega X_\alpha(\omega)$ and the corresponding operators in the interaction picture are given by 
\begin{equation}
  \tilde{X}_\alpha(\omega,t) \equiv e^{i\Hil t} X_\alpha(\omega) e^{-i \Hil t} = e^{-i\omega t} X_\alpha(\omega).
\end{equation}
Using these operators, we can write the system-bath interaction term in the interaction picture as
\begin{equation}
  \HcbBT(t)\!= \!\!\sum_{\alpha,\omega}\!\left[
    \sigma_-^\alpha(\omega)e^{-i\omega t} \! \otimes \! \tilde{B}_\alpha^\dagger (t) \! + \! \sigma_+^\alpha(\omega)e^{+i\omega t} \! \otimes \! \tilde{B}_\alpha(t)\right].
\end{equation}
Substituting the eigenoperator expansion into Eq.~\eqref{eq:redfield0} and evaluating the bath trace using the correlation functions \eqref{eq:C+}--\eqref{eq:C-}, we can write the Redfield equation as,
\begin{eqnarray}
  \frac{d\tilde{\rho}(t)}{dt} \! & = & \!  \sum_\alpha \sum_{\omega,\omega'}  \Bigl\{
    e^{i(\omega'-\omega)t} \widetilde{\Cc}_\alpha^-(\omega)\! \Big[ \sigma_-^\alpha (\omega) \tilde{\rho}(t) \sigma_+^\alpha (\omega')  \nonumber \\
    && \qquad \qquad   -\; \sigma_+^\alpha(\omega')\sigma_-^\alpha(\omega) \tilde{\rho}(t) \Big] \nonumber\\
  && +\,e^{i(\omega'-\omega)t}\widetilde{\Cc}_\alpha^+(\omega)\!\Big[ \sigma_+^\alpha (\omega') \tilde{\rho}(t) \sigma_-^\alpha(\omega) \nonumber \\
  && \qquad \qquad  - \;\sigma_-^\alpha (\omega) \sigma_+^\alpha (\omega') \tilde{\rho}(t) \Big] \Bigr\} +\text{h.c.},
  \label{eq:redfield2}
\end{eqnarray}
where the one-sided Fourier transforms of the bath correlation functions, $\widetilde{\Cc}_\alpha^\pm(\omega)=\int_0^\infty d\tau\,e^{i\omega\tau}\Cc_\alpha^\pm(\tau)$ are given by
\begin{eqnarray}
  \widetilde{\Cc}_\alpha^+(\omega) & = & \frac{\Gamma_\alpha(\omega)}{2} \nF(\omega,T_\alpha) +\,i\,\frac{\mathcal{H}_\alpha^+(\omega)}{2} \label{c+1} \\
  \widetilde{\Cc}_\alpha^-(\omega)   & = &\frac{\Gamma_\alpha(\omega)}{2} [1-\nF(\omega,T_\alpha)] +\,i\,\frac{\mathcal{H}_\alpha^-(\omega)}{2} \label{c-1}
\end{eqnarray}
with 
\[ \Gamma_\alpha(\omega)=2\pi J_\alpha(\omega) = 2\pi(\lambda^\alpha(\omega))^2 D_\alpha(\omega) \sim 2 \pi (\lambda^\alpha)^2 D_\alpha,\]
being the bare Fermi's-golden-rule decay rate at frequency $\omega$ (cf.\ Eq.~\eqref{eq:spectral_density}), and $\mathcal{H}_\alpha^\pm(\omega)$ being the principal-value part that appears as a Lamb-shift term in the master equation for the charger-battery system. Notice that under the Markov approximation $\Gamma_\alpha(\omega) \sim \Gamma_\alpha$. 

Now we can make the third approximation, namely the secular approximation. The Redfield equation, \eqref{eq:redfield2}, contains terms oscillating at frequencies $\omega'-\omega$ for all pairs of eigen-frequencies. When converted back to the
Schr\"{o}dinger picture, these give rise to terms oscillating with frequencies, $|\omega'-\omega|$. In order to obtain a GKSL type master equation for the charger-battery system from the Redfield equation, we have to coarse grain over a timescale that is fast compared to the timescale of free evolution of the qubits but slow compared to the relaxation rates of the qubits due to the baths. This coarse-graining, or averaging over a timestep $\Delta t$ as $\int_0^{\Delta t} \, dt$ of the Redfield equation removes all terms for which $\omega \neq \omega'$ and this constitutes the Secular or Rotating Wave approximation. 

Applying the coarse graining, secular approximation and substituting Eqs.~\eqref{c+1}--\eqref{c-1} into the Redfield equation we obtain, 
\begin{eqnarray}
  \frac{d\tilde{\rho}(t)}{dt} \!\! & = \!\! & \!  \sum_{\alpha, \omega}   \Bigl\{
    \Gamma_\alpha^-(\omega)\! \Big[ \sigma_-^\alpha (\omega) \tilde{\rho}(t) \sigma_+^\alpha (\omega)  \nonumber \\
    && \qquad    -\; \frac{1}{2} \sigma_+^\alpha(\omega)\sigma_-^\alpha(\omega) \tilde{\rho}(t) - \frac{1}{2} \tilde{\rho}(t) \sigma_+^\alpha(\omega)\sigma_-^\alpha(\omega)  \Big] \nonumber\\
  && \! +\Gamma_\alpha^+(\omega)\!\Big[ \sigma_+^\alpha (\omega) \tilde{\rho}(t) \sigma_-^\alpha(\omega) \nonumber \\
  && \qquad   - \;\frac{1}{2}\sigma_-^\alpha (\omega) \sigma_+^\alpha (\omega) \tilde{\rho}(t) - \frac{1}{2}\sigma_-^\alpha (\omega) \tilde{\rho}(t)\sigma_+^\alpha (\omega)  \Big] \nonumber \\
  && \!\! - i \frac{{\mathcal H}_\alpha^-(\omega)}{2} \! \bigl[ \sigma_+^\alpha \sigma_-^\alpha, \tilde{\rho}(t) \bigr] \!\! - \!i \frac{{\mathcal H}_\alpha^+(\omega)}{2} \! \bigl[ \sigma_-^\alpha \sigma_+^\alpha, \tilde{\rho}(t) \bigr] \!\!\Bigr\} , \qquad
  \label{eq:redfield3}
\end{eqnarray}
with
\begin{eqnarray}
\Gamma_\alpha^+(\omega) & = &  \Gamma_\alpha(\omega) \nF(\omega,T_\alpha) \sim \Gamma_\alpha \nF(\omega,T_\alpha), \nonumber \\
\Gamma_\alpha^-(\omega) & = &  \Gamma_\alpha(\omega) [1-\nF(\omega,T_\alpha)] \sim \Gamma_\alpha [1-\nF(\omega,T_\alpha)] . \qquad
\end{eqnarray}

we find that the non-unitary effect of the reservoirs on the qubits is captured by dissipators of the form, 
\[ {\mathcal L}_\alpha = \sum_{\omega} \Big( \Gamma_\alpha^-(\omega) {\mathcal D}[\sigma_-^\alpha(\omega)] + \Gamma_\alpha^+ (\omega) {\mathcal D}[\sigma_+^\alpha(\omega)] \Big), \]
where $\mathcal{D}[X]\bullet \equiv X\bullet X^\dagger - \frac{1}{2}(X^\dagger X \bullet + \bullet X^\dagger X)$ is the standard jump operator GKSL master equation.  

The rates $\Gamma_\alpha^\pm(\omega_s)$ govern the two competing processes by which bath $\alpha$ acts on each of the two qubits.  $\Gamma_\alpha^-(\omega_s)$ is the rate at which the qubit relaxes and loses energy $\omega$. The associated jump operator is $\sigma_-^\alpha(\omega)$, which is a lowering operator for the system. This process requires the bath to absorb the emitted quantum, which is possible only if a bath mode at frequency $\omega_s$ is unoccupied. Hence the factor $[1-\nF(\omega_s,T_\alpha)]$ in $\Gamma_\alpha^-(\omega_s)$. Similarly, $\Gamma_\alpha^+(\omega)$ is the rate at which the qubit is excited by the bath through the associated raising operator  $\sigma_+^\alpha(\omega)$. This process requires the bath to donate a quantum, which is possible only if a bath mode at frequency $\omega$ is occupied. Hence the factor $\nF(\omega_s,T_\alpha)$ in $\Gamma_\alpha^+(\omega)$. 

These rates satisfy the KMS (Kubo-Martin-Schwinger) detailed balance condition,
\begin{equation}
  \frac{\Gamma_\alpha^+(\omega)}{\Gamma_\alpha^-(\omega)}=e^{-\omega/T_\alpha}.
  \label{eq:kms}
\end{equation}
The Boltzmann factor $e^{-\omega_s/T_\alpha}$ quantifies the asymmetry between excitation and relaxation: the bath excites the qubit exponentially less readily than it relaxes it, by precisely the factor required for thermal equilibrium. The KMS condition guarantees that the steady state of each bath dissipator alone is the correct Gibbs state $e^{-\HilT/T_\alpha}/Z_\alpha$.

The reservoirs also generate a Lamb-shift correction to $\Hil$ given by
\begin{equation}
  H_\text{LS}^\alpha=\sum_{\omega}
  \frac{\mathcal{H}_\alpha(\omega)}{2}\sigma_+^\alpha(\omega)\sigma_-^\alpha(\omega),
  \label{eq:HLS_alpha}
\end{equation}
where  ${\mathcal H}_\alpha (\omega) = {\mathcal H}_\alpha^+(\omega) + {\mathcal H}_\alpha^-(\omega)$. In arriving at this expression for $H_\text{LS}^\alpha$ we have ignored a c-number term proportional to $[\sigma_+^\alpha(\omega), \sigma_-^\alpha(\omega)]$. 

\subsubsection{The bath frequencies \label{bathfreq}}

We need to identify the Fourier components of $\sigma_\pm^\alpha$ and the corresponding frequencies. For the charger qubit the non-zero matrix elements of the lowering operator $\sigma_-^c$ with respect to the eigenstates of $\Hil$ are,
\begin{eqnarray}
  \bra{\psi_{11}}\sgm{c}\ppl=\cos\tfrac{\theta}{2}, \;\;
  \bra{\psi_{11}}\sgm{c}\pmi=-\sin\tfrac{\theta}{2},\nonumber \\
  \bra{\psi_{01}}\sgm{c}\pee=\cos\tfrac{\theta}{2}, \;\;
  \bra{\psi_{10}}\sgm{c}\pee=-\sin\tfrac{\theta}{2}.\nonumber
\end{eqnarray}
However these four matrix elements correspond to only two different frequencies, namely, 
\begin{eqnarray}
   \omega_+= \Lambda_{01}-\Lambda_{11} = \Lambda_{00}-\Lambda_{10}= \frac{\omega_c + \omega_b}{2} + \Omega, \nonumber \\
   \omega_-= \Lambda_{10}-\Lambda_{11} = \Lambda_{00}-\Lambda_{01}= \frac{\omega_c + \omega_b}{2} - \Omega.
   \label{frequencies2}
\end{eqnarray}
The secular approximation keeps cross terms between transitions at the same frequency and so the two matrix elements at each frequency combine into composite rank-2 jump operators given by,
\begin{eqnarray}
  \sigma_-^c(\omega_+)&=\cos\tfrac{\theta}{2} \, \pgg \bra{\psi_{01}}  - \sin\tfrac{\theta}{2} \,\pmi \bra{\psi_{00}},\nonumber \\
  \sigma_-^c(\omega_-)&=-\sin \tfrac{\theta}{2} \, \pgg \bra{\psi_{10}} +\cos \tfrac{\theta}{2}\, \ppl \bra{\psi_{00}},\nonumber
\end{eqnarray}
and $\sgm{c}=\sigma_-^c(\omega_+)+\sigma_-^c(\omega_-)$.

The same two frequencies appear for the battery qubit also and the corresponding rank-2 jump operators for the battery qubit are given by, 
\begin{eqnarray}
  \sigma_-^b(\omega_+)&=\sin\tfrac{\theta}{2}\,\pgg\bra{\psi_{01}} +\cos \tfrac{\theta}{2} \, \pmi \bra{\psi_{00}}, \nonumber \\
  \sigma_-^b(\omega_-)&=\cos\tfrac{\theta}{2} \, \pgg \bra{\psi_{10}} +\sin\tfrac{\theta}{2} \, \ppl \bra{\psi_{00}}, \nonumber
\end{eqnarray}
with $\sgm{b}=\sigma_-^b(\omega_+)+\sigma_-^b(\omega_-)$.

The imaginary (principal-value) part of the one-sided Fourier transform at each of the frequencies $\omega_+$ and $\omega_-$ generates a coherent Lamb shift Hamiltonian as give in Eq.~\eqref{eq:HLS_alpha} with 
\begin{equation}
  \mathcal{H}_\alpha(\omega_\pm)=\frac{\mathcal{P}} {\pi}\!\int \frac{ \Gamma_\alpha(\omega')}{\omega'-\omega_\pm}\,d\omega'
  = \frac{2}{\pi}\mathcal{P}\!\int\frac{J_\alpha(\omega')}{\omega'-\omega_\pm}\,d\omega'.
  \label{eq:lamb_bath}
\end{equation}
Since $\sigma_-^\alpha(\omega_\pm)$ are rank-2 operators with off-diagonal elements in the dressed basis, the products $\sigma_+^\alpha(\omega_\pm)\sigma_-^\alpha(\omega_\pm)$ have both diagonal and off-diagonal contributions, renormalizing the effective coupling
$\gamma$ as well as the level splittings.

In the Schrodinger picture, the dynamics of the charger and battery system under the influence of the Fermionic baths coupled to them is then described by the master equation, 
\begin{equation}
  \frac{d\rho}{dt}\!=\!-i\!\left[\Hil \! + \! H_\text{LS}^c \! + \! H_\text{LS}^b,\;\rho \right] \! + \! \Lc_c[\rho] \! + \! \Lc_b[\rho].
\label{eq:full_me}
\end{equation}

Apart from the weak, continuous measurement introduced later, $\Hil$, the Lamb-shift terms and the dissipators $\mathcal{L}_\alpha$ furnish a complete description of our
charger-battery system. If we are to compare quantum and classical batteries on the same footing, it is important to ensure that the total energy that can be stored in either case is the same. The key figure of merit for comparing classical and quantum batteries is therefore the \textit{charging rate},
defined as
\begin{equation}
    P(t) = \frac{d}{dt}\langle H_{\rm b} \rangle_t,
    \label{chargingrate}
\end{equation}
where $H_{\rm b} = E_{\rm b}n_{\rm b} = E_{\rm b}(\openone + \sigma_z^{\rm b})/2$, with $n_{\rm b} \in \{0,1\}$, is the operator corresponding to the total energy stored in the battery. The expectation value in Eq.~\eqref{chargingrate} is computed with respect to the
instantaneous state $\rho_{\rm cb}(t)$ of the charger-battery pair as $\langle H_{\rm b}\rangle_t = {\rm Tr}[(H_{\rm b} \otimes \openone_{\rm c})\rho_{\rm cb}(t)]$.
The charging rate then decomposes into coherent and dissipative terms,
\begin{equation}
    P_{\text{coh}}(t)  = \mathrm{Tr}\!\big(i[\Hil,\,H_{\rm b} \otimes \openone_{\rm c}] \rho_{\rm cb}(t)\big),
\end{equation}
which represents the instantaneous energy flow due to coherent interaction between the battery qubit and the charger qubit, and
\begin{align}
P_{\text{diss}}(t) = \sum_{\alpha={\rm c}, {\rm b}} \Big(
    &\mathrm{Tr}[L_\alpha^\dagger (H_{\rm b} \otimes \openone_{\rm c}) L_\alpha \rho_{\rm cb}(t)]
    \nonumber\\
    &- \tfrac{1}{2}\mathrm{Tr}[\{L_\alpha^\dagger L_\alpha,\,  H_{\rm b} \otimes \openone_{\rm c}\}\rho_{\rm cb}(t)] \Big),
\end{align}
describing energy changes due to reservoirs coupled to the two qubits. This
form for the dissipative part of the charging rate is particularly useful
because the introduction of weak, continuous measurements on one or both
qubits can be modelled as the addition of appropriate dissipative terms to
$P_{\rm diss}$.

\section{Measurements Using a Quantum Point Contact}
\label{sec:MeasurementDissipator}

\begin{figure}[htbp]
    \centering
    \includegraphics[width=\linewidth]{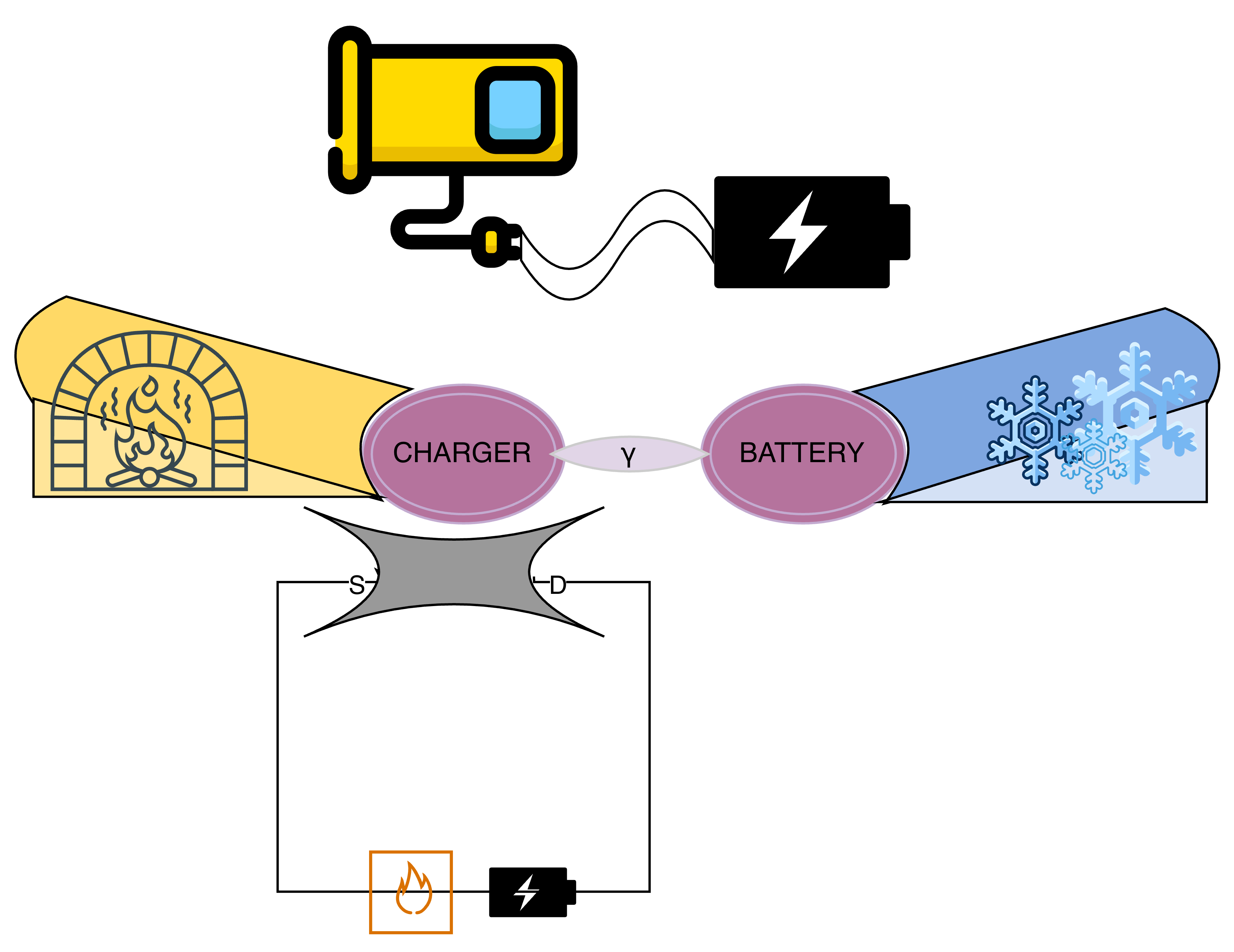}
    \caption{Charger-battery setup with the charger capacitively coupled to a
    QPC device performing weak, continuous measurement.}
    \label{fig:single_qpc_setup}
\end{figure}

We now introduce weak, continuous measurements of one or both of the qubits in our charger-battery system with the aim of exploring how such measurements can modify the charging rate of the battery qubit. We first discuss the case where only one of the qubits is measured, with the measurement performed by coupling a quantum point contact (QPC) to the qubit. In the case of double quantum-dot based charge qubits, we assume that the QPC is kept in close proximity to one of the two dots. Conduction through the QPC is quantized, and we assume that the QPC is calibrated so that it operates at the edge of one of the steps in its quantized conductivity in the absence of an electron in the neighboring dot. If an electron enters the dot, it changes the electrostatic potential at the QPC, shifting it to a lower conductivity step. The presence
of an electron in the dot that is close to the QPC is thus detectable as a change in conductivity, and in effect, the QPC measures the $\sigma_z$ operator on the coupled qubit.

In this section, we provide the key elements of the microscopic derivation of the measurement-induced dynamics for our quantum system coupled to a QPC detector. We restrict our discussion to the essential aspects of the
derivation of the jump operator that describes the back-action of the QPC on the measured qubit, following the detailed discussion given in Refs.~\cite{Goan2001,Korotkov1999,Wiseman2009,Li2013,Hussein2014,Ouyang2010}. The Lamb shift induced on the qubits due to the interaction with the QPC is also computed.

The QPC consists of an electrostatically gated constriction with a ``Source''~(S) reservoir of electrons on one side and a ``Drain''~(D) reservoir on the other. Electrons move from S to D driven either by a temperature gradient or by a potential difference between the two reservoirs. In this section we assume that the QPC is coupled only to one of the two qubits and does not interact with the reservoirs coupled to either qubit. The total Hamiltonian for the two qubits, the QPC, and the source and drain (ignoring for the time being, the baths whose effects we have already computed) is 
\begin{equation}
H = H_0 + H_{\rm meas},
\end{equation}
where $H_0 = \Hil + H_S + H_D$ with
\begin{equation}
H_S = \sum_q \epsilon_q^S s_q^\dagger s_q, \quad
H_D = \sum_p \epsilon_p^D d_p^\dagger d_p,
\end{equation}
and $\Hil$ given in Eq.~\eqref{eq:hamiltonian}. Here, $s_q$ and $d_p$ are Fermionic annihilation operators associated with the source and drain reservoirs of the QPC, respectively.

The interaction between the QPC and the qubit coupled to it is
\begin{equation}
H_{\rm meas} = \sum_{qp} (g_{qp} + \chi_{qp} M)(s_q^\dagger d_p + d_p^\dagger s_q),
\label{eq:interaction}
\end{equation}
where $M$ is the measurement operator acting on the qubit, $g_{qp}$ is the bare tunneling amplitude across the QPC, and $\chi_{qp}$ represents the measurement coupling strength. Assuming the QPC couples to the charger qubit, the measurement operator is
\begin{equation}
M = \sigma_z^{\rm c} = \sigma_z^{\rm c}\otimes \openone_b,
\end{equation}
We rewrite the interaction Hamiltonian that controls the QPC measurement as
\begin{equation}
H_\text{meas}=S_g\otimes{F}_g+S_\chi\otimes{F}_\chi,
  \quad S_g=\one,\quad S_\chi=\sgz{c},
\end{equation}
where ${F}_g$ and ${F}_\chi$ are Hermitian bath operators carrying amplitudes $g_{qp}$ and $\chi_{qp}$ respectively.

\subsection{Effect of the QPC \label{QPCmaster}}

The same three-step procedure used for the baths is applied to the QPC coupling. The QPC is assumed to be a wide-band reservoir in a non-equilibrium steady state (maintained by the bias $eV$), with correlation functions decaying on a timescale $\tau_\text{QPC}\ll 1/ \Gamma_\text{meas}$. The interaction picture with respect to $\Hil + H_S + H_D$ brings in the time dependence of the system operators $S_g=\one$ and $S_\chi=\sgz{c}$. Since $S_g=\one$ is always stationary, all non-trivial structure comes from the eigen-operator decomposition of $S_\chi=\sgz{c}$

Since $[\sgz{c},\Hil]\neq0$, $\sgz{c}$ is not stationary in the interaction picture and following the earlier procedure we compute its matrix elements with respect to the eigenstates of $\Hil$. Corresponding to $\omega=0$, we have the following eigenstates:
\begin{eqnarray}
  \bra{\psi_{00}}\sgz{c}\pee=+1, & \;\; \bra{\psi_{11}}\sgz{c}\pgg=-1, \nonumber \\
  \bra{\psi_{01}}\sgz{c}\ppl=+\cos\theta,& \;\; \bra{\psi_{10}}\sgz{c}\pmi=-\cos\theta. \nonumber
\end{eqnarray}
Corresponding to $\omega =\pm 2 \Omega$ we have the off-diagonal matrix elements, 
\begin{equation}
  \bra{\psi_{01}} \sgz{c} \pmi = \bra{\psi_{10}} \sgz{c} \ppl = -\sin\theta.
\end{equation}
The full eigen-operator decomposition of $\sigma_z^c$ is $ \sgz{c}=\sgz{c}(0) + \sgz{c}(2\Omega) + \sgz{c}(-2\Omega)$ with the zero-frequency eigen-operator
\begin{eqnarray}
  \sgz{c}(0)& = & \pee \bra{\psi_{00}} - \pgg \bra{\psi_{11}} \nonumber \\
    && \qquad +\cos\theta \! \left( \ppl \bra{\psi_{01}} - \pmi \bra{\psi_{10}} \right), \label{QPCzeroOp}
\end{eqnarray}
and finite-frequency eigenoperators
\begin{eqnarray}
  \sgz{c}(+2 \Omega)&=&-\sin \theta\,\ppl \bra{\psi_{10}}, \nonumber \\
  \sgz{c}(-2\Omega)&=&-\sin\theta\, \pmi\bra{\psi_{01}} = \sgz{c \dagger}(2\Omega). \label{QPCOmegaOps}
\end{eqnarray}
In the interaction picture, $\sgz{c}$ has the time dependence
\begin{equation}
  \tilde{\sigma}_z^c(t)=\sgz{c}(0)+ e^{-2i\Omega t} \sgz{c}(2\Omega) +e^{+2i \Omega t} \sgz{c}(-2\Omega).
\end{equation}

The Born approximation in this case allows use to write $\tilde{\varrho}(t) = \tilde{\rho}_{\rm cb} \otimes \rho_{QPC}$ and following the same steps as before we find that the relevant two-time correlation functions of the QPC that we have to evaluate have the form $\Cc_{\alpha\beta}(\tau) =\langle F_\alpha(t) F_\beta (t-\tau) \rangle_\text{QPC}$ for $\alpha,\beta\in\{g,\chi\}$. However since $S_g=\openone$, the double commutator in Eq.~\eqref{eq:born} that contains only $S_g$ vanishes and we need not evaluate $\langle F_g(t) F_g (t-\tau) \rangle_\text{QPC}$. We now anticipate the Markov approximation and assume that the QPC coupling spectrum is wide with $g_{qp} \sim g$ and $\chi_{qp}\sim \chi$. The two time correlation functions of the QPC now have the form, 
\[  \Cc_{\alpha\beta}(\tau) \! = \! \lambda_{\alpha\beta}\cdot\Cc(\tau), \, \lambda_{gg} \! = \! g^2, \, \lambda_{\chi\chi} \! = \! \chi^2, \, \lambda_{g\chi} \! = \! \lambda_{\chi g} \! = \! g\chi.\]
As before, their one-sided Fourier transforms split into real (dissipative) and imaginary (unitary/dispersive) parts,
\begin{equation}
  \tilde{\Cc}_{\alpha\beta}(\omega)=\frac{S_{\alpha\beta}(\omega)}{2}
  +\frac{i\mathcal{H}_{\alpha\beta}(\omega)}{2}.  \label{QPCpower}
\end{equation}

The QPC can be driven either by a temperature difference between the source and the drain or a voltage bias, which is equivalent to a difference in chemical potential between the two leads or a combination of the two. The source and the drain are characterized by their respective Fermi-Dirac distributions given by, 
 \begin{equation}
  \nS(\epsilon)=\frac{1}{e^{(\epsilon-\mu_{\rm S})/T_{\rm S}}+1},\;
  \nD(\epsilon)=\frac{1}{e^{(\epsilon-\mu_{\rm D})/T_{\rm D}}+1},
\end{equation}
Where $T_S$ and $T_D$ ($T_S > T_D$) are the temperatures of the source and the drain while $\mu_S$ and $\mu_D$ are their respective chemical potentials. The QPC is  driven out of equilibrium by both a temperature difference $\Delta T = T_S - T_D = T_{meas}$ and a chemical potential difference, that is equivalent to a voltage difference of $V$ such that $|eV|=\mu_D-\mu_S$. The current through the QPC therefore has two  contributions: a voltage-driven (drift) component and a thermally driven (diffusion) component.

Once the secular approximation is also applied assuming that the dissipation due to the interaction with the QPC is much slower than the typical dynamical time scale of the qubits, the dissipative dynamics due to the QPC is described by the jump operators $\sgz{c}(0)$ and $\sgz{c}(\pm 2\Omega)$ given in Eqs,~\eqref{QPCzeroOp} and \eqref{QPCOmegaOps}. However the decay rates are controlled by the the noise power $S_{\alpha\beta}$ that appears in Eq.~\eqref{QPCpower}. The noise power, in turn, comes from the two time correlation functions of the QPC which, in the interaction picture, is built from pairs of source
and drain Fermionic operators. A forward tunneling event moves an electron from the drain mode at energy $\epsilon_p$ to the source mode at energy $\epsilon_q$, contributing a phase factor $e^{i(\epsilon_q-\epsilon_p)\tau}$. Notice that while the electron is being transferred from the drain to source, the current is from source to drain. In the wide band limit we introduce $\epsilon\equiv\epsilon_q-\epsilon_p$ as the net energy transferred, and the two time correlation function becomes becomes
\begin{eqnarray}
  \mathcal{C}(\tau) & = & D_{\rm S} D_{\rm D}   \int_{-\infty}^\infty d\epsilon\, \bigl[ \nS (\epsilon) (1 - \nD(\epsilon)) \, e^{-i\epsilon\tau} \nonumber \\
  && \qquad \qquad  \quad  + \; \nD(\epsilon)(1-\nS(\epsilon))\,e^{i\epsilon\tau}\bigr],
  \label{eq:C_QPC_tau}
\end{eqnarray}
where $D_{\rm S}$ and $D_{\rm D}$ are the density of states for the source and drain respectively. On taking the one-sided Fourier transform to get $\tilde{C}(\omega)$, the real part gives $S(\omega)/2$ and the imaginary part gives $i\mathcal{H}(\omega)/2$, with
\begin{equation}
 S(\omega) \! = \! 2\pi D_{\rm S} D_{\rm D} \! \bigl[\nS (\omega)[1 \! - \! \nD(\omega)] \! + \! \nD(-\omega)[1 \! - \! \nS(-\omega)] \bigr].
  \label{eq:S_evaluated}
\end{equation}
This term is the analogue of $\Gamma_\alpha(\omega)=2\pi J_\alpha(\omega)$ for the Fermionic baths considered earlier.  The noise power $S_{\alpha\beta}(\omega)$ for each $\alpha$ and $\beta$ is obtained by multiplying
$S(\omega)$ by $\lambda_{\alpha\beta}$. It is worth noting here that a positive potential bias between the source and drain ($\mu_{\rm S} > \mu_{\rm D}$) drives electrons from the drain to the source due to the negative charge of the electrons while a positive thermal bias ($T_{\rm S} > T_{\rm D}$) drives electrons from source to drain. At the  {\em Seebeck voltage}, the two effects cancel each other out and the net current through the QPC is zero. However even at zero current $S(0) > 0$ and the corresponding dephasing rate $\Gamma_{\rm QPC}(0)$ is also nonzero due to fluctuations in electron tunneling even without a net drift.  

\subsubsection{QPC Dissipators and Lamb shifts \label{QPCdiss}}

From Eqs.~\eqref{QPCzeroOp} and \eqref{QPCOmegaOps} there are three relevant frequencies at which the QPC affects the charger-battery system, namely $\omega =0$ and $\omega= \pm 2\Omega$. The term $\lambda_{gg} S(0)$ does not have any effect on the charger-battery system since the corresponding system operator is $S_g = \openone$. The term containing $\lambda_{g\chi}S(0)$ also vanishes since the Redfield equation contains the double commutator $[\tilde{S}_g(t), \, \tilde{S}_\chi(t-\tau) \, \rho ]
  = [ \openone_{\rm c},\, \tilde{\sigma}_z^c(t-\tau)\, \tilde{\rho}(t) ]=0$. The $\lambda_{\chi g}S(0)$ term, on the other hand, is non zero and produces only a Lamb-shift term, $H_{\rm LS}^{g\chi} = \lambda_{\chi g} S(0)\sigma_z^c(0)$, with $\sigma_z^c(0)$ given in Eq.~\eqref{QPCzeroOp}. The Lamb shift renormalizes both $\Omega$ and the coupling between the qubits, $\gamma$. 

  In the $\chi-\chi$ sector we get three disspipators. Corresponding to $\omega = 0$ and the jump operator $\sigma_z^c(0)$ we have the dissipator, 
  \begin{equation}
  \Lc_\text{meas}^{(0)}[\rho]=\Gamma^0_{\rm meas} \! \left[ \sgz{c}(0) \rho \,\sgz{c}(0)\,-\tfrac{1}{2}\{(\sgz{c}(0))^2, \rho \} \right],
  \end{equation}
with dephasing rate given by $\Gamma^0_{\rm meas}=\pi\chi^2 D_{\rm S} D_{\rm D} S_{\rm FF}(0)$. 
Corresponding to $\omega=-2\Omega$ and the jump operator $\sgz{c}(-2\Omega)=-\sin \theta \, \pmi\bra{\psi_{01}}$ that drives the transition $\ppl \to \pmi$ at rate $\Gamma^-_{\rm meas}=\sin^2\theta \pi \chi^2 D_{\rm S} D_{\rm D}  S_{\rm FF}(-2\Omega)$ we have the dissipator, 
\begin{eqnarray}
  \Lc_\text{meas}^{(-)}[\rho] & = & \Gamma^-_{\rm meas} \!\bigl[ \pmi \bra{\psi_{01}} \rho \, \ppl \bra{\psi_{10}} \nonumber \\
  && \qquad - \tfrac{1}{2} \{ \ppl \bra{\psi_{01}}, \rho\} \bigr].
\end{eqnarray}
Similarly, corresponding to $\omega=+2\Omega$ and jump operator $-\sin\theta \, \ppl \bra{\psi_{10}}$, driving $\pmi\to\ppl$ at rate $\Gamma^+_{\rm meas} = \sin^2 \theta  \pi \chi^2 D_{\rm S} D_{\rm D}  S_{\rm FF}(+2\Omega)$, we have the dissipator, 
\begin{eqnarray}
  \Lc_\text{meas}^{(+)}[\rho] & = & \Gamma^+_{\rm meas} \! \bigl[ \ppl \bra{\psi_{10}} \rho \pmi\bra{\psi_{01}} \nonumber \\
  && \qquad - \tfrac{1}{2} \{ \pmi \bra{\psi_{10}}, \rho\} \bigr].
\end{eqnarray}
The imaginary parts of $\hat{\Cc}_{\chi\chi}(\pm2 \Omega)$ generate:
\begin{equation}
  H_\text{LS}^{\chi\chi} \! = \! \frac{\sin^2 \theta}{2}\!\left[ \mathcal{H}(2\Omega)\pmi \! \bra{\psi_{10}}  \! + \! \mathcal{H}(-2\Omega) \ppl \! \bra{\psi_{01}}\right].
\end{equation}
This renormalises the $\ppl$--$\pmi$ splitting, proportional to $\chi^2\sin^2\theta$ and is vanishing at $\gamma=0$. The net effect is that introduction of the QPC adds two Lamb shift terms $H_{\rm LS}^{g\chi}$ and $H_{\rm LS}^{\chi \chi}$ as well as three dissipators, $\Lc_\text{meas}^{(0)}$ and $\Lc_\text{meas}^{(\pm)}$ to Eq.~\eqref{eq:full_me}.

\subsection{Numerical investigations}
\label{sec:numerical}

\begin{figure}
    \centering
     \includegraphics[width=0.95\linewidth]{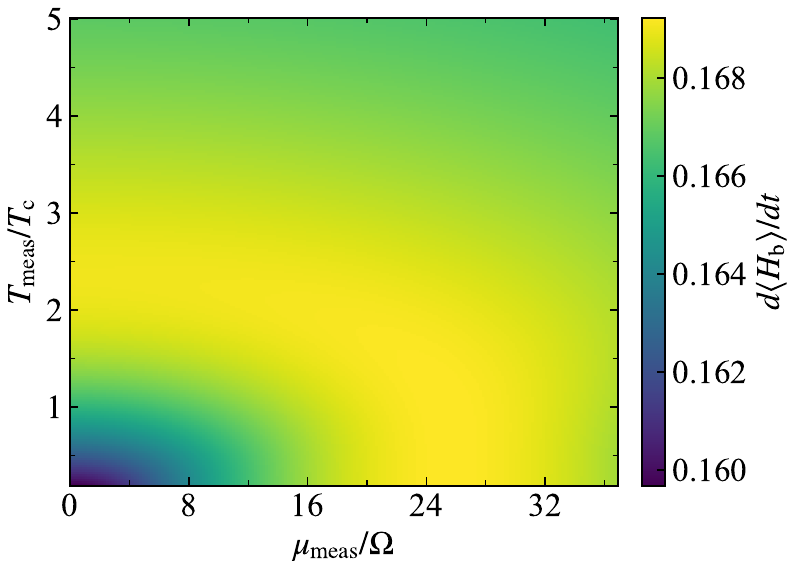} \hfill
    \includegraphics[width=0.95\linewidth]{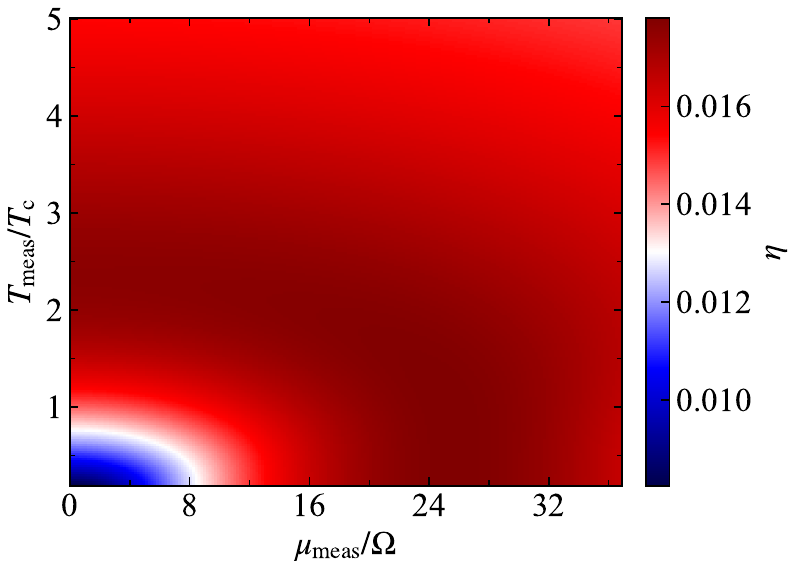}
        \caption{  The figure on top shows the charging rate of the quantum battery as a function of the measurement strength parameters, $T_{\rm meas}/T_{\rm c}$ and $\mu_{\rm meas}/\Omega$. The Difference in the charging rate between the measured and unmeasured configurations is shown in the bottom figure. We see that the charging rate initially increases and has a maximum for certain values of the measurement parameters. For higher measurement strengths, the charging rate is suppressed.}
    \label{fig:single_qpc_results}
\end{figure}

Before analyzing the measurement-induced effects, we specify the initial operating point of the charger-battery system. Following the framework established in Ref.~\cite{PhysRevResearch.7.013151}, we prepare the system in a non-equilibrium steady state (NESS) governed by the bare charger-battery master equation in the absence of any measurement apparatus. The NESS remains invariant under time evolution described by Eq.~\eqref{eq:full_me}.   The system parameters are chosen as $\omega_{\rm c} = 4g$, $\omega_{\rm b} = 3g$, $T_{\rm c} = 6g$, $T_{\rm b} = 2g$, $\gamma = 0.212g$, and the bath qubit couplings which are approximately constant under the Markov approximation are taken to be $(\lambda^{\rm c} )^2  =(\lambda^{\rm b} )^2= 0.1g$, where $g$ ($ \sim \mathrm{kHz}$) sets the natural energy scale. This NESS serves as the initial state. We then switch on the QPC measurement channel and evolve the system under the full master equation that incorporates the measurement back-action, driving the charger-battery system to a new steady state whose properties we now examine. 

We study the behavior of the charger-battery system as a function of strengths of the thermal gradient and potential difference that drives electrons across the QPC. The thermal gradient is characterized by $T_{\rm meas} = T_{\rm S} - T_{\rm D}$ However as noted earlier, a positive potential gradient, $\mu_{\rm S} > \mu_{\rm D}$ drives electrons across the QPC in a direction opposite to that of $T_{\rm meas}$. We are however not interested in this competition between the two effects and instead, we focus on the the measurement back action that arises from driving electrons across the QPC from the source to the drain. For this reason, we take $\mu_{\rm meas} = \mu_{\rm D} - \mu_{\rm S}$ as the second parameter we vary while studying the effect of the measurements on the charger-battery system. In particular, our independent variables are the scaled driving parameters $T_{\rm meas}/T_{\rm c}$ and $\mu_{\rm meas}/\Omega$. We scale the measurement temperature with respect to the reservoir temperature $T_{\rm c}$ and not with respect to the typical energy scale $\Omega$ because the relevant non equilibrium transport processes are controlled by thermal occupation functions and temperature gradients between reservoirs~\cite{Kosloff2014,Myers2024}.

The main effect we are interested in is the measurement back-action on the charger qubit. Continuous monitoring by the QPC detector induces stochastic, non-unitary dynamics on the charger, effectively disturbing its dynamics  and driving a redistribution of populations among the energy levels of the qubit. This measurement-induced dephasing channel can inject energy into the charger by promoting transitions to excited states. This is a process that can be understood as the QPC acting as an entropic and energetic resource for the composite quantum charger-battery system. The Lamb-shifts also create changes in the energy levels and couplings between the charge and the battery. A natural question that arises is whether it is fair to include the Lamb shifts $H_{\rm LS}^{g\chi}$ and $H_{\rm LS}^{\chi \chi}$ when assessing how the QPC measurements enhance or suppress the charging rate. The Lamb shift terms which are proportional to the imaginary part of the one-sided Fourier transform of the bath correlation functions correspond to off-shell, virtual processes. It involves no exchange of quanta or information between the QPC and the battery-charger system and it only a dressing of the system energy levels by the presence of the bath. The jump operators, on the other had, represent genuine back action and are related directly to the information extracted by the QPC current. We therefore assume that the Lamb shifts are absorbed into the bare frequencies and couplings and in practice, for the simulations, which compare the cases with and without measurements, we set the Lamb shifts effectively to zero. 

To quantify the effect of the QPC dissipators, we examine the back-action energy $\Delta E_{\mathrm{BA}}$ imparted to the charger by the measurement apparatus. $\Delta E_{\mathrm{BA}}$ exhibits a sign change as a function of the QPC operating parameters. In the weak-measurement regime which is characterized by low thermal bias $T_{\rm meas}/T_{\rm c}$ and a small  potential bias $\mu_{\rm meas}/\Omega$, the back-action energy is negative, indicating that the detector is insufficiently driven to inject energy into the system and its effect is to remove energy from the charger.  As the measurement strength increases, the back-action energy crosses into the positive regime and the QPC drives substantial population inversion in the charger, enabling a net average energy gain through the measurement process alone. In this regime, the act of measurement itself constitutes a thermodynamic resource.

We find that both measurement-induced energy removal and injection into the charger translates into  enhanced charging rates for the quantum battery relative to the case where the charger is not measured.   To isolate the measurement-induced contribution, we define the enhancement in charging rate of the battery as as
\begin{equation}
    \eta = \frac{d \langle H_{\rm b} \rangle_{\mathrm{meas}}}{dt} - \frac{d \langle H_{\rm b} \rangle_{\mathrm{unmeas}}}{dt},
    \label{eq:enhancement}
\end{equation}
where ${d \langle H_{\rm b} \rangle_{\mathrm{meas}}}/{dt}$ denotes the charging rate in the presence of the QPC detector. From Fig.\ref{fig:single_qpc_results} we observe a non-monotonic dependence of the enhancement of the charging rate on the measurement strength. Beyond a characteristic threshold the enhancement begins to diminish, with a maximum in the middle for certain ranges of $T_{\rm meas}/T_{\rm c}$ and $\mu_{\rm meas}/\Omega$. This behavior delineates an optimal operating regime which forms an annular region in the $(T_{\rm meas}/T_{\rm c} , \, \mu_{\rm meas}/\Omega)$ parameter space, within which the measurement-induced charging enhancement remains near its maximum value. From a practical standpoint, this identifies a robust plateau where the quantum battery can be operated to extract the maximum benefit from measurement-assisted charging without the need for excessive fine-tuning the detector parameters. 

\begin{figure}[!htb]
    \centering
    \includegraphics[width=0.95\linewidth]{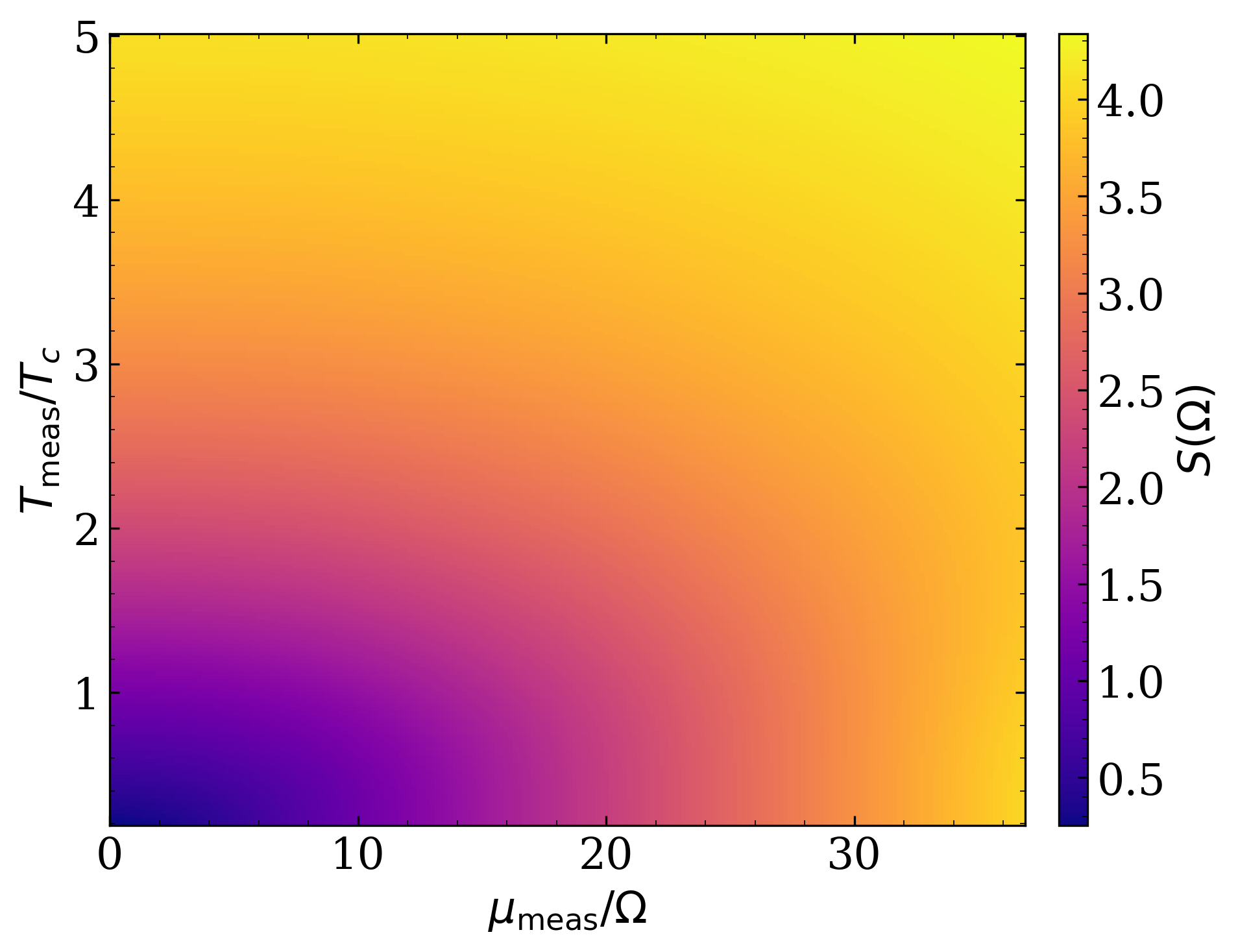}
    \caption{ Dependence of $S(\Omega)$  on $\mu_{\rm meas}$ and $T_{\rm meas}$. We see that the noise power of the detector is a monotonically increasing function of both measurement parameters.}
    \label{fig:gamma_dep}
\end{figure}

It is interesting to note that while the charging rate shows a maximum, the effect of the QPC on the charger as measured by the decay and dephasing rates, $\Gamma^0_{\rm meas}$ and $\Gamma^{\pm}_{\rm meas}$ are monotonically increasing functions of both $\mu_{\rm meas}$ and $T_{\rm meas}$ as shown in fig.~\ref{fig:gamma_dep}. This follows directly from the structure of the noise power $S(\omega)$ increasing $\mu_{\rm meas}$ raises the source occupation $n_{\rm S}(\omega)$, enlarging the bias window available for electron tunneling, while increasing $T_{\rm meas}$ broadens both Fermi functions and populates higher-energy states, both effects only adding to the net current through the QPC.  As a consequence, the measurement back-action on the system grows monotonically with the thermal and potential-based driving of the detector, and the parameter space of interest is bounded not by the rate functions themselves but by the non-monotonic response of the battery charging rate to increasing measurement strength.

\section{Measurements using two independent and series coupled QPCs}
\label{sec:correlated}

We now consider an extended measurement device in which two QPCs are placed in series along a single conduction channel,
\[   \text{Source}\;(S)
  \xrightarrow{\;\text{QPC}_1\;}
  \text{Channel}\;(C)
  \xrightarrow{\;\text{QPC}_2\;}
  \text{Drain}\;(D), \]
with the charger qubit  coupled to QPC$_1$ and the battery qubit coupled to QPC$_2$. The source and drain are common to the whole device, and the channel $C$ is the intermediate region between the two QPCs. The three Fermionic reservoirs are described by the respective Hamiltonians,
\begin{equation}
  H_\text{res}=\sum_q\epsilon^{\rm S}_q s_q^\dagger s_q
  +\sum_k\epsilon^{\rm C}_k c_k^\dagger c_k
  +\sum_p\epsilon^{\rm D}_p d_p^\dagger d_p,
\end{equation}
where $s_q$, $c_k$, $d_p$ are the annihilation operators for the source, the channel, and the drain modes respectively, satisfying the usual Fermionic anti-commutation relations. The source and drain are maintained at chemical potentials $\mu_{\rm S}$ and $\mu_{\rm D}$  and temperatures $T_{\rm S}$ and $T_{\rm D}$ respectively as before. The channel can be of two types. It can be considered as a region in which the electrons passing through effectively thermalize with no residual coherence that is carried from the first QPC to the second. In this case we effectively have two independent QPCs with their own respective temperature and potential gradients. In other words, the two QPCs will be performing independent measurements of the charger and battery qubits respectively. The second option is to consider the channel as a short quantum wire through with the electrons pass through without loss of coherence. 

The qubit-dependent tunneling amplitudes at each QPC are structured exactly as in the single-QPC case, but now the two system operators are $\sgz{c}$ (charger, at QPC$_1$) and $\sgz{b}$ (battery, at QPC$_2$) with the interaction described by, 
\begin{eqnarray} 
H_{\text{meas},\rm{c}} & = & \sum_{qk} \! \left(g_{qk}^{\rm c} + \chi_{qk}^{\rm c} \sgz{c} \right)  \!\left(s_q^\dagger c_k+c_k^\dagger s_q\right), \nonumber \\
H_{\text{meas},\rm{b}} & = & \sum_{kp}\!\left(g_{kp}^{\rm b} + \chi_{kp}^{\rm b} \sgz{b}\right)   \!\left(c_k^\dagger d_p+d_p^\dagger c_k\right).
  \label{eq:Hint1}
\end{eqnarray}
We can decompose each interaction as:
\begin{eqnarray*}
  H_{\text{meas},{\rm c}} & = & S_g^{\rm c} \otimes \hat{F}_g^{\rm c} + S_\chi^{\rm c} \otimes \hat{F}_\chi^{\rm c},
  \quad  S_g^{\rm c} =\openone,\; S_\chi^{\rm c}=\sgz{c},   \\
  H_{\text{meas},{\rm b}} & = & S_g^{\rm b}\otimes\hat{F}_g^{\rm b}+S_\chi^{\rm b}\otimes\hat{F}_\chi^{\rm b},
  \quad S_g^{\rm b} = \openone,\; S_\chi^{\rm b}=\sgz{b},  
\end{eqnarray*}
where the bath operators are
\begin{eqnarray*}
  \hat{F}_{g}^{\rm c}(t)&=\sum_{qk}g_{qk}^{\rm c}\!
  \left(s_q^\dagger c_k\,e^{i(\epsilon_q-\epsilon_k)t}+\text{h.c.}\right),\\
  \hat{F}_{\chi}^{\rm c}(t)&=\sum_{qk}\chi_{qk}^{\rm c}\!
  \left(s_q^\dagger c_k\,e^{i(\epsilon_q-\epsilon_k)t}+\text{h.c.}\right),\\
  \hat{F}_{g}^{\rm b}(t)&=\sum_{kp}g_{kp}^{\rm b}\!
  \left(c_k^\dagger d_p\,e^{i(\epsilon_k-\epsilon_p)t}+\text{h.c.}\right), \\
  \hat{F}_{\chi}^{\rm b}(t)&=\sum_{kp}\chi_{kp}^{\rm b}\!
  \left(c_k^\dagger d_p\,e^{i(\epsilon_k-\epsilon_p)t}+\text{h.c.}\right).
\end{eqnarray*}
The system operators are now $S_g^{\rm c}=\one$, $S_\chi^{\rm c}=\sgz{c}$ at the first QPC and $S_g^{\rm b}=\one$, $S_\chi^{\rm b}=\sgz{b}$ at the second QPC. This introduces two new features relative to the single-QPC case. Two distinct measurement operators: $\sgz{c}$ and $\sgz{b}$, whose eigen operator decompositions in the dressed basis are different and the decomposition fo $\sgz{b}$ must be computed analogously to that for $\sgz{c}$. Secondly, there are now cross terms between the two QPCs. The Born-Markov equation now contains pairs $(\alpha,\beta)$ where $\alpha$ belongs to QPC$_{\rm c}$ and $\beta$ belongs to QPC$_{\rm b}$, mediated by the shared channel $C$. These cross terms generate non-local, charger-battery correlated, contributions to the master equation when the channel is coherent.

\begin{figure}[!htb]
    \centering
    \includegraphics[width=0.9\linewidth]{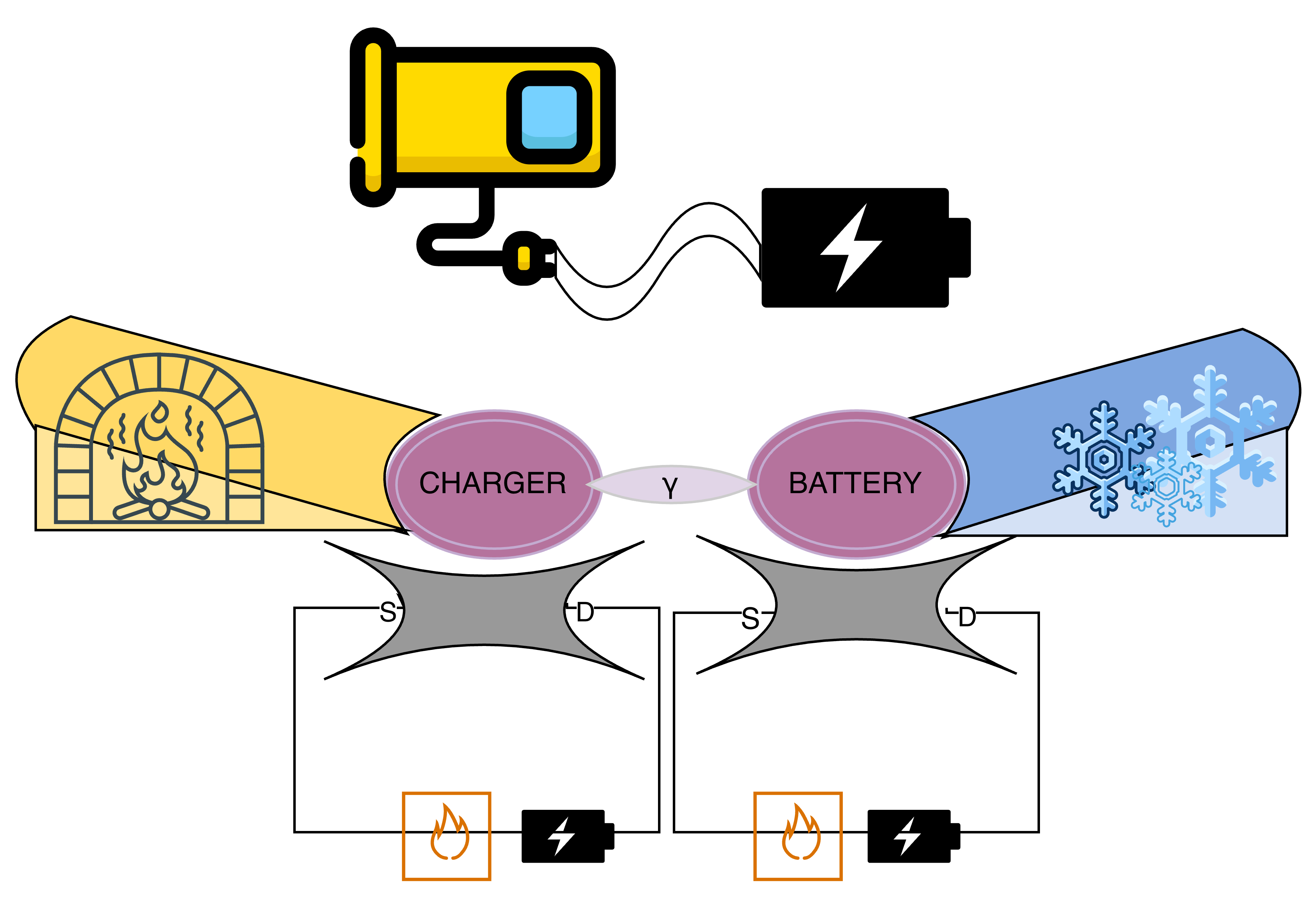}
    \hfill
    \includegraphics[width=0.95\linewidth]{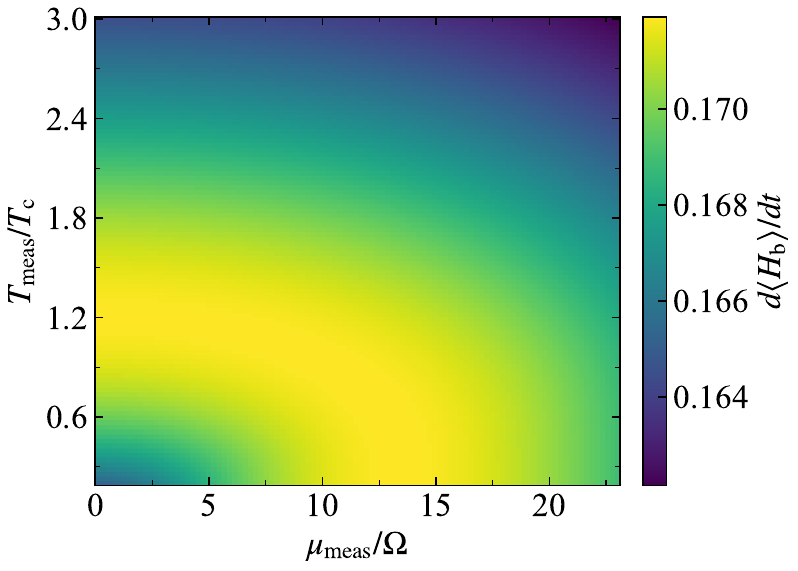}
    \hfill
    \includegraphics[width=0.95\linewidth]{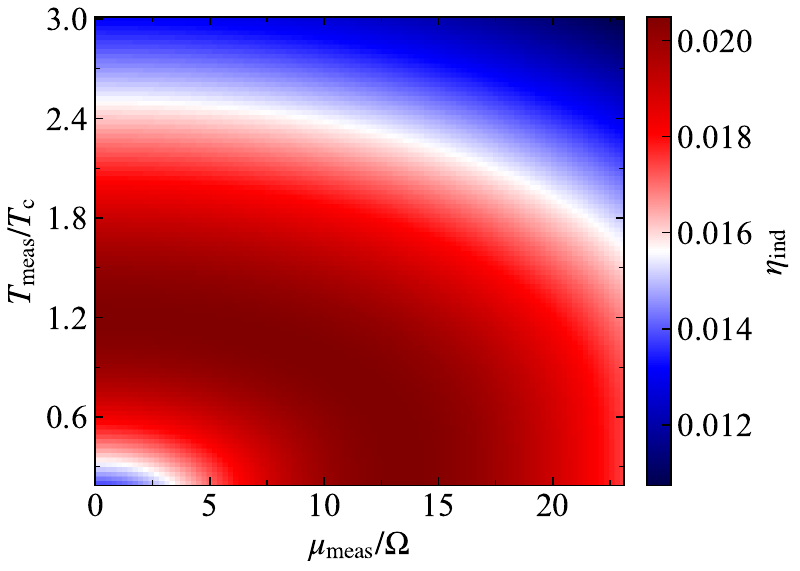}
    \caption{Independent two-QPC measurement configuration.
    The top panel shows the schematic of the two  QPC configuration in which both the charger and battery qubits are independently and continuously monitored. The middle panel shows the  charging rate of the quantum battery as a function of $T_{\rm meas}/T_c$ and $\mu_{\rm meas}/\Omega$ while the bottom panel show the enhancement in charging rate, $\eta_{\mathrm{ind}}$ relative to the unmeasured baseline.}
    \label{fig:two_qpc_results}
\end{figure}

As far as the effect of QPC$_{\rm c}$ is concerned, the jump operators are still given by Eqs.~\eqref{QPCzeroOp} and \eqref{QPCOmegaOps}. The noise power $S(\omega)$ is modified with $D_{\rm C}$ and $n_{\rm C}(\omega)$ replacing $D_{\rm D}$ and $n_{\rm D}(\omega)$ respectively since the drain for the first QPC is the channel. The decoherence rates $\Gamma^{0, \pm}_{{\rm meas}, {\rm c}}$ are also modified in the same manner. The effect of the QPC connected to the battery qubit is captured by the jump operators, 
\begin{eqnarray}
  \sgz{b}(0)& = & \pee \bra{\psi_{00}} - \pgg \bra{\psi_{11}} \nonumber \\
    && \qquad -\cos\theta \! \left( \ppl \bra{\psi_{01}} - \pmi \bra{\psi_{10}} \right), \label{QPCbzeroOp}
\end{eqnarray}
and
\begin{eqnarray}
  \sgz{b}(+2 \Omega)&=&\sin \theta\,\ppl \bra{\psi_{10}}, \nonumber \\
  \sgz{b}(-2\Omega)&=&\sin\theta\, \pmi\bra{\psi_{01}} = \sgz{b \dagger}(2\Omega). \label{QPCbOmegaOps}
\end{eqnarray}
The noise power is given by 
\begin{equation}
 S(\omega) \! = \! 2\pi D_{\rm C} D_{\rm D} \! \bigl[n_{\rm C} (\omega)[1 \! - \! \nD(\omega)] \! + \! \nD(-\omega)[1 \! - \! n_{\rm C}(-\omega)] \bigr],
  \label{eq:S_evaluatedb}
\end{equation}
and the decay rates are
\begin{eqnarray*}
    \Gamma^0_{{\rm meas}, {\rm b}} & = &  \pi (\chi^{\rm b})^2 D_{\rm C} D_{\rm D} S_{\rm FF}(0), \\
    \Gamma^{\pm}_{{\rm meas}, {\rm b}} &= &\sin^2\theta \pi (\chi^{\rm b})^2 D_{\rm C} D_{\rm D}  S_{\rm FF}(\pm 2\Omega).
\end{eqnarray*}
The net effect of the second QPC is to add three more dissipators, ${\mathcal L}_{{\rm meas}, {\rm b}}^{(0, \pm)}$ to the master equation. This QPC will also add Lamb shifts analogous to QPC$_{\rm c}$ but as noted before we ignore these terms in our numerical simulations and they can be absorbed as renormalization of the bare frequencies and couplings. The effect of QPC$_{\rm b}$ is the new effect that comes in when the channel $C$ is incoherent. Equivalently one can think of this scenario as a case where there are two independent QPCs connected to the charger and battery qubits respectively. The charging rate and its enhancement in the case where we treat the two QPCs as independent and driven by identical temperature and potential gradients is shown in Fig.~\ref{fig:two_qpc_results}. We see that while the behavior is qualitatively similar to that of the single QPC case, the optimal values of $T_{\rm meas}$ and $\mu_{\rm meas}$ are smaller by roughly a factor of 2 as expected since we are doubling the measurement back-action by having two QPCs instead of one. 

In addition to the optimal values of $T_{\rm meas}$ and $\mu_{\rm meas}$ becoming smaller with two independent QPCs, we see that the the numerical values of the charging rate and enhancement at the optimal points are larger with two QPCs. This means that having both the qubits observed increases the charging rate even though the resources in terms of the total energy given to the electrons traversing the QPCs is comparable with the single QPC case. 

\subsection{Correlated Measurements}
\label{sec:results_correlated}

When the channel is a short coherent wire, then the channel modes $c_k$ are do not constitute a thermal reservoir and instead, they are a set of quantum states through which electrons propagate coherently.  We can integrate these modes out perturbatively to obtain an effective source-to-drain Hamiltonian in which the channel no longer appears explicitly. Treating $H_{{\rm meas}, {\rm c/b}}$ from Eq.~\eqref{eq:Hint1} as perturbations on $H_{\rm res}$, we see that a source to drain effective coupling appears only at second order in perturbation with the channel modes acting as intermediate (virtual) states. The effective Hamiltonian acting on the subspace consisting of the source, the drain and the two qubits can be written as a T-matrix expression, 
\begin{equation}
  H_\text{eff} \! = \! H_{\text{meas},{\rm c}} \! \frac{1}{\nu-H_{\rm C}} \! H_{\text{meas}, {\rm b}}
  + H_{\text{meas},{\rm b}} \! \frac{1}{\nu-H_{\rm C} } \! H_{\text{meas}, {\rm c}},
  \label{eq:Tmatrix}
\end{equation}
where $H_{\rm C}=\sum_k \epsilon_k c_k^\dagger c_k$ is the channel Hamiltonian and $\nu$ is the energy of the electron that enters the channel, which, in turn, is set by the source temperature and potential. Let us now consider a process in which an electron initially in the source mode $q$ with the channel modes unoccupied, $|q,0\rangle$ transitions through the channel into a drain mode $p$ with the corresponding state denoted by $|p,0\rangle$. We have, 
\begin{eqnarray}
    \langle p,0|H_{\rm eff}|q,0\rangle & = & \langle p,0| H_{\text{meas},{\rm b}} \! \frac{1}{\nu-H_{\rm C} } \! H_{\text{meas}, {\rm c}}|q,0\rangle, \nonumber \\
    & = & \sum_k \frac{(g_{qk}^{\rm c}+\chi_{qk}^{\rm c}\sgz{c})(g_{kp}^{\rm b}+\chi_{kp}^{\rm b}\sgz{b})}  {\epsilon_q-\epsilon_k},
  \label{eq:forward}
\end{eqnarray}
where we have used the fact that the electron energy $\epsilon$ is almost the same as the energy of the source mode, $\epsilon_q$. If we assume a wide band spectrum for the reservoirs so as to justify the Markov approximation with $g_{qk}^{\rm c} \approx g^{\rm c}$, $\chi_{qk}^{\rm c} \approx \chi^{\rm c}$, $g_{kp}^{\rm b} \approx g^{\rm b}$, $\chi_{kp}^{\rm b} \approx \chi^{\rm b}$, the energy denominator factorizes out and we can evaluated the (retarded) energy denominator as
\begin{eqnarray*}
  \sum_k \frac{1} {\epsilon_q-\epsilon_k+i0^+} & \longrightarrow &  \int d \epsilon_k\,\frac{D_{\rm C}}{\epsilon_q - \epsilon_k+i0^+}  \\ &= & D_{\rm C}\,\mathcal{P}\! \int \frac{d \epsilon_k}{\epsilon_q - \epsilon_k} -i \pi D_{\rm C}.
\end{eqnarray*}
The principal value part of the integral leads to a Lamb shift term which can be absorbed into the relevant couplings of the two qubit Hamiltonian while the imaginary term contributes to the coherent tunneling rate across the channel from source to drain. This terms contribute to the tunneling amplitude through the T-matrix element, ${\mathcal T}_{qp} = -i\pi D_{\rm c}t_{\rm eff}^{\rm{qp}}$, where,
\begin{equation}
    t_\text{eff}^{qp} = g^{\rm c}g^{\rm b}\,\openone
  + g^{\rm c}\chi^{\rm b}\,\sgz{\rm b}
  + \chi^{\rm c}g^{\rm b}\,\sgz{\rm c}
  + \chi^{\rm c}\chi^{\rm b}\,\sgz{\rm c} \sgz{\rm b}.
\end{equation}
This leads to the effective Hamiltonian, 
\begin{equation}
    H_\text{eff}=\sum_{qp}t_\text{eff}^{qp}\!\left(s_q^\dagger d_p+d_p^\dagger s_q\right). 
\end{equation}

There are four types of system operators in $H_{\rm eff}$, namely $\openone$, $\sgz{c}$, $\sgz{b}$ and $\sgz{c} \otimes \sgz{b}$. The Lamb-shifts, jump operators and decay rates corresponding to the first three operators are identical to the case of the two independent QPCs. The new operator that appears in the case of correlated measurements is  $\sgz{c} \otimes \sgz{b}$. This operator commutes with the interaction term $\gamma\!\left(\sgp{c}\sgm{b}+\sgm{c}\sgp{b}\right)$ that appears in the two qubit Hamiltonian $H_{\rm cb}$ and hence it commutes with $H_{\rm cb}$ also. This means that  $\sgz{c} \otimes \sgz{b}$ is diagonal in the eigenbasis of $H_{\rm cb}$ and so in the interaction picture, it has only as zero frequency component, 
\[  \sgz{\rm cb} (0) \! = \!  \pee \! \bra{\psi_{00}} + \pgg \! \bra{\psi_{11}}  -\ppl \! \bra{\psi_{01}} - \pmi \! \bra{\psi_{10}}. \]
Corresponding to this jump operator, we obtain using the same steps as before, the decay rate, 
\[ \Gamma_0^{zz}=(\chi^{\rm c}\chi^{\rm b })^2\,S^{12}(0), \]
where $ S^{12}(\omega)= \bar{D}_{\rm C}^2 S(\omega)$, with $S(\omega)$ given in Eq.~\eqref{eq:S_evaluated} and $\bar{D}_{\rm C} = \int d\epsilon_k\, D_{\rm C}[1-n_{\rm C}(\epsilon_k)]$.

We see that the coherent measurements  gives rise to additional dissipative contributions in the master equation that couple the
decoherence channels of the charger and battery subsystems. These
cross-correlations qualitatively modify the non-equilibrium steady state and, consequently, the energy transport pathways within the composite system. 

\begin{figure}[!htb]
    \centering
    \includegraphics[width=0.9\linewidth]{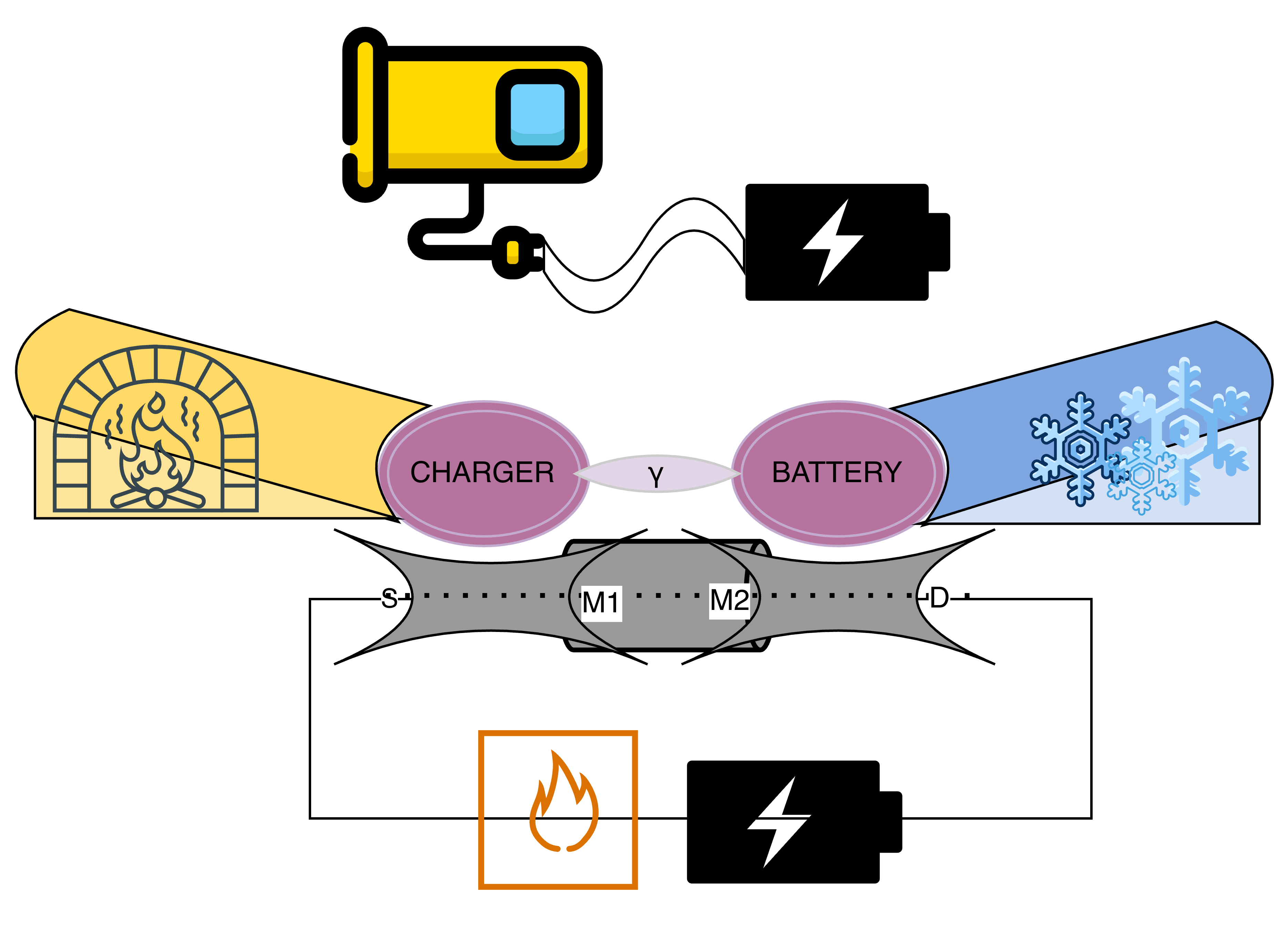}\\
    \includegraphics[width=0.85\linewidth]{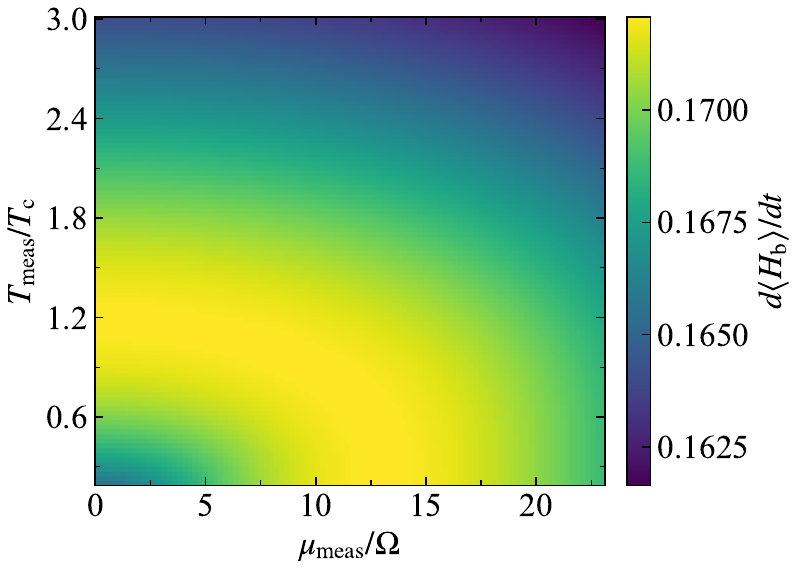}\\
    \includegraphics[width=0.85\linewidth]{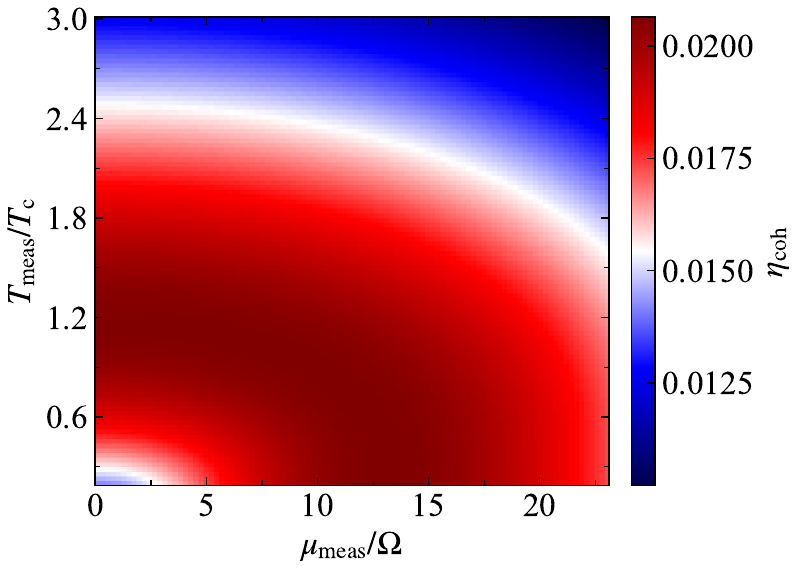}
    \caption{ Coherent two-QPC measurement configuration. The top panel shows a schematic diagram of the series-coupled QPC geometry. The middle panel shows the steady-state charging rate of the quantum battery as a function of the measurement parameters $T_{\rm meas}/T_{\rm c}$ and $\mu_{\rm meas}/\Omega$. The bottom panel shows the charging rate enhancement $\eta_{\mathrm{coh}} =
    \dot{E}_B^{(\mathrm{coh})} - \dot{E}_B^{ ( \mathrm{unmeas} ) }$ relative to the unmeasured baseline, demonstrating improvement over the single-QPC case both in terms of charging rate as well as resource requirements.}
    \label{fig:series_qpc_results}
\end{figure}

While the charging rate and its enhancement in Fig.~\ref{fig:series_qpc_results} looks qualitatively similar to those in Fig.~\ref{fig:two_qpc_results} for two independent QPCs, the comparison of the two hides a significant point. When we have two independent QPCs, the resources required to drive the measurements are captured by the thermal and potential gradients across each of the two reservoirs. We found previously that these gradients are roughly half of that for the single QPC at the optimal points indicating that in the two cases the resource requirement is roughly the same as in the required gradients becoming half when the number of QPCs is doubled. There was a modest increase in the maximum attainable charging rate with two QPCs indicating that indeed having the two QPCs is advantageous. With the two QPCs connected in series through a coherent channel, we see that the charging rate and its enhancement is comparable to the two independent QPC case but the resource requirement is genuinely half of that the single QPC case because a single temperature and potential gradient is driving both the QPCs and these gradients are half of that for the single QPC case at the optimal points. 

There is another, sub-leading contribution to the charging rate enhancement in the coherent case. This comes from the jump operator that arises from the $\sigma_z^{\rm c} \otimes \sigma_z^{\rm b}$ term. This contribution is small since the decay rate is proportional to $(\chi^{\rm c} \chi^{\rm b})^2$ which is fourth order in $\chi$ as opposed to the other rates which are second order in $\chi$. This small contribution is not noticeable in Figs.~\ref{fig:two_qpc_results} and \ref{fig:series_qpc_results} but we can see it from the numerically obtained maximum value of the charging rate in each case which is found to be 0.1692 for the single QPC, 0.1719 for the two independence QPCs and 0.1721 for the coherent case respectively.

The correlated dissipator term that appears when the two QPC measurement is coherent appears to engineer a preferential energy transfer channel between the charger and the battery. Unlike the single-QPC and independent QPC cases, where the measurement back-action acts locally on the qubits. The energy flow to the battery is normally mediated by the coherent coupling $\gamma$ is augmented by a dissipation-assisted transport pathway characterized by $\Gamma_0^{zz}$ that directly facilitates energy
redistribution across the charger-battery partition. This in turn appears to maximize the fraction of measurement back-action energy that is converted into useful battery charging. In the case of two independent QPCs, double the back action has to be supplied to observe a behavior comparable to the coherent case.  In comparison with both the unmeasured baseline, the single-QPC
configuration and the two independent QPC case, the coherent measurement scheme establishes the most favorable trade-off between charging rate enhancement and measurement
resource expenditure, underscoring the role of measurement correlations as a key design principle for measurement-powered quantum batteries.

\section{Discussion and Conclusion}
\label{sec:discussion}

We now consolidate the results obtained across the three measurement configurations we considered and discuss their physical implications for measurement-assisted quantum battery charging. Our main finding is captured in the enhancement hierarchy $ \eta < \eta_{\mathrm{ind}} < \eta_{\mathrm{coh}}$,
which orders the three measurement configurations, the single QPC, independent two-QPC, and coherent measurement schem in terms of the magnitude of charging rate increase relative to the unmeasured baseline and the minimal resources consumed to achieve it. Table~\ref{tab:qpc_comparison} summarizes the optimal
operating parameters and corresponding enhancements for each configuration against a common unmeasured baseline.
\begin{table}[htb]
\centering
\begin{tabular}{lccccc}
\hline \hline \\[-2 mm]
 Configuration   &  $T_{\rm meas}/T_{\rm c}$ & $ \mu_{ \rm meas}/\Omega$ & $P(t)$ & $\eta$ (\%)  \\[1 mm] \hline \\[-3 mm]
Single QPC  & 0.6583 & 25.551 & 0.1692 & 11.76 \\[1 mm]
Two QPCs    & 0.5960 & 12.820 & 0.1719 & 13.54 \\[1 mm]
Coherent  & 0.5394 & 12.589 & 0.1721 & 13.64 \\[1 mm]
\hline\hline \\[-2 mm]
\end{tabular}
\caption{Comparison of the three QPC measurement configurations at their respective optimal operating points. The common unmeasured baseline is ${P}^{(0)}(t)=0.1514$.}
\label{tab:qpc_comparison}
\end{table}

We find that the coherent two-QPC configuration achieves the highest charging rate enhancement of $13.64\%$, attained at the lowest measurement apparatus driving among all three
configurations. This demonstrates that the cross-correlated dissipator, which is unique to the coherent measurement configuration, opens a dissipation-assisted energy transfer channel that steers energy flow along the thermodynamically optimal trajectory from charger to battery. The correlated measurement thus achieves a dual advantage namely,  superior enhancement at reduced resource expenditure.

The independent two-QPC configuration yields a comparable but
marginally lower enhancement of $13.54\%$, but the resource expenditure in terms of the temperature and potential gradients required is almost double that of the coherent case as well as the single QPC case. Although the introduction of a second measurement channel acting locally on the battery provides additional back-action-induced population redistribution, the absence of the cross-correlated dissipator  prevents this
configuration from matching the efficiency of the coherent scheme. The near-degeneracy of the two enhancements which are separated by only by around $0.1\%$ difference highlights that the dominant contribution to the enhancement arises from monitoring both subsystems. However, the correlations provide a significant advantage in terms of the resources required. We also see that local measurement back-action on the charger alone, while sufficient to demonstrate the principle of measurement-assisted charging, is inherently less effective and efficient compared to  the configurations that engage both subsystems.

\begin{acknowledgments} 
A.~S.~acknowledges funding support from the National Quantum Mission through the Foundation for Quantum Computing Innovation, an initiative of the Department of Science and Technology, Government of India
\end{acknowledgments}

\bibliographystyle{apsrev4-2}
\bibliography{references}

\end{document}